\documentclass[twocolumn]{aastex63}

\usepackage{amsmath}
\usepackage[cm]{sfmath}

\usepackage{rotating}

\newcommand{\mearth}{$\mathrm{M}_\oplus$}

\received{June 19, 2020}
\revised{August 6, 2020}
\accepted{August 7, 2020}

\submitjournal{PSJ}

\shorttitle{Colliding in the shadows of giants}
\shortauthors{Carter et al.}

\graphicspath{{./}{figures/}}

\begin{document}

\title{Colliding in the shadows of giants: Planetesimal collisions during the growth and migration of gas giants}

\correspondingauthor{Philip J. Carter}
\email{pjcarter@ucdavis.edu}

\author[0000-0001-5065-4625]{Philip J. Carter}
\affiliation{Department of Earth and Planetary Sciences, University of California, Davis, One Shields Avenue, Davis, CA 95616, USA}

\author[0000-0001-9606-1593]{Sarah T. Stewart}
\affiliation{Department of Earth and Planetary Sciences, University of California, Davis, One Shields Avenue, Davis, CA 95616, USA}

\begin{abstract}

Giant planet migration is an important phenomenon in the evolution of planetary systems. Recent works have shown that giant planet growth and migration can shape the asteroid belt, but these works have not considered interactions between planetesimals. We have calculated the evolution of planetesimal disks, including planetesimal-planetesimal collisions, during gas giant growth and migration. The numbers, locations, and impact velocities of these collisions depend on the specific growth and migration path. We find that giant planet growth alone has little effect on impact velocities, and most of the planetesimals scattered by growing giants do not undergo collisions with each other during the growth period. In contrast, we find that giant planet migration induces large numbers of high velocity collisions between planetesimals. These impacts have sufficient velocities to cause shock-induced vaporization for both water ice and silicate components of planetesimals, and to cause catastrophic disruption of the bodies. New bodies may form from impact debris. Collisional evolution reduces the efficiency of planetesimal implantation into the asteroid belt via giant planet growth and migration. A small fraction of the largest planetesimals implanted into the asteroid belt would have been processed via collisions. We identify important consequences of planetesimal collisions that have not been considered in planet accretion models. The prevalence of high velocity collisions during giant planet migration, and their potential links to the properties of meteorites, and the formation of chondrules, makes impact vaporization a critically important phenomenon. The consequences of vaporizing planetesimal constituents require further detailed study. New collision outcome models for impacts within the nebula, and models for new planetesimal formation are needed.

\end{abstract}

%\keywords{planetesimals -- migration -- collisions -- planet formation}

%%%%%%%%%%%%%%%%%%%%% INTRO %%%%%%%%%%%%%%%%%%%%%% 

\section{Introduction}\label{s:intro}

Our understanding of planet formation and planetary compositions, is based on the remnants of the accretion process that we can observe in our solar system and in exoplanet systems today. Besides the planets themselves, our solar system contains numerous asteroids from which our inventory of meteorites is derived. The so-called primitive, chondritic meteorites were once thought to represent the building blocks of the planets. However, it is now clear that chondrite parent bodies and planets formed contemporaneously \citep[e.g.][]{Dauphas11,Scott14}. The relationship between meteorites and planets is unclear. 

There are several major outstanding questions regarding the origins and histories of meteorites that have been delivered to Earth. Perhaps the most confounding are the processes that control the timing and formation of planetesimals ({meter to kilometer }sized bodies) and their components, such as calcium-aluminum-rich inclusions (CAIs) and chondrules \citep[e.g.][]{Desch12,ConnollyJones16}. Chondrules are previously molten, mm-sized grains found in chondrites. Chondrules formed during the first few Myr of solar system history and, as such, are considered an important component of planetary evolution, but their uncertain origin is a topic of debate \citep[e.g.][]{ConnollyJones16}. The bodies from which chondrites originated did not form in isolation; to understand these enigmatic meteorites we must consider the effects of their formation environment.

During the time period in which the meteorite parent bodies were forming, the nebular gas was still present in the disk and the giant planets were still accreting \citep[e.g.][]{Kleine05,Becker15,Sanborn18}. Perturbations from giant planets can substantially excite planetesimal orbits. These excited orbits cause bow shocks as the fast-moving planetesimals travel through the nebular gas \citep{Hood09}, and increase planetesimal impact velocities \citep{Dobinson16,Shoichi19}. The gas giants are believed to have played a key role in sculpting the asteroid belt \citep[e.g.][]{Walsh11,Johnson16,Raymond17,Deienno18,Bryson20}. While the dynamical aspects of the giant planets influence on planetesimals have recently been studied in detail \citep[e.g.][]{Turrini12,Raymond17}, the consequences of collisional processing during the growth and possible migration of Jupiter (and Saturn), when meteorite parent bodies were forming, have generally been ignored. The overlap in formation times, and the uncertain timing of giant planet formation and the dissipation of the nebular gas disk, make understanding the relationship between meteorites and planets a complex problem.

Collisions between planetary building blocks are essential to the formation of planets. Giant impacts between planetary embryos have been studied in great detail \citep[e.g.][]{Asphaug06,Marcus09,Genda12,Burger18,Carter20,Gabriel20}, as have collisions between small planetesimals \citep[e.g.][]{Stewart09,Jutzi10,Leinhardt12,Movshovitz16}. However, these studies have generally ignored the presence of the nebular gas, which can change collision outcomes \citep{Hood09,StewartLPSC19a,Davies19}. Collision outcomes depend critically on the impact velocity as well as the masses and compositions of the colliding bodies. 
Collisional processing has been shown to be important in the growth of terrestrial bodies in the inner solar system \citep[e.g.][]{Carter15,Dwyer15,Davies20}, and may play an important role in setting the properties of meteorites, but collisions in the vicinity of the giant planets and in the outer solar system, have not been studied in detail.

\citet{Raymond17} showed that the growth and migration of the gas giants causes substantial excitation of planetesimal orbits, scattering some of those that formed close to Jupiter into the asteroid belt region. However, these authors did not consider the collisions that may have occurred during this period. It has commonly been assumed that aerodynamic drag from the nebular gas would rapidly damp planetesimal orbits, and impacts between planetesimals would have been low-velocity, merging events. 
However, during gas giant growth and migration especially, the strong gravitational influence of the giant planets can drive planetesimals to attain large eccentricities despite the drag from the nebula. High eccentricity planetesimals can have high velocities relative to circular Keplerian orbits, leading to high impact velocities.

Impacts at several km\,s$^{-1}$ have been shown to eject `jets' of melt droplets that have been suggested as a chondrule formation mechanism \citep{Johnson15, Wakita17}. \citet{Davies19b} and \citet{Davies20} found that collisions would begin to cause vaporization of ice and rock at impact velocities of $\sim$1\,km\,s$^{-1}$ and $\sim$8\,km\,s$^{-1}$ respectively, velocities that can readily be achieved when planetesimal orbits are excited by the giant planets \citep{Carter15,Johnson16,Hin17,Davies20}. Impact-induced vaporization may also play a role in the formation of chondrules and chondrites \citep{Hood09,Carter19,StewartLPSC19b}.

Several previous studies have attempted to quantify the distribution of impact velocities during the growth or migration of giant planets. \citet{Turrini12} showed impact velocities reach up to 10\,km\,s$^{-1}$ in the asteroid belt due to the growth of Jupiter, but these impact velocities were based on probabilistic calculations, not directly resolved collisions. \citet{Johnson16} showed that the migration of Jupiter in the Grand Tack model could induce impact velocities greater than 20\,km\,s$^{-1}$ in the asteroid belt, however, they considered only impacts onto embryos, or only the asteroid belt region in isolation.

Here, for the first time, we examine in detail the locations and frequencies of directly resolved planetesimal-planetesimal collisions that occur in the vicinity of growing and migrating giant planets. In this work we focus on the short duration `exciting' events of growth and migration rather than long term evolution and growth. We examine the impact velocity distributions and collision rates obtained from $N$-body simulations, and discuss the implications for the meteorite record.

%%%%%%%%%%%%%%%%%%%% METHODS %%%%%%%%%%%%%%%%%%%%%

\section{Numerical methods}

We carried out sets of simulations designed to capture the dynamics and collisions within planetesimal disks during the rapid accretion and migration of gas giant planets. We used a modified version of the parallelized $N$-body code PKDGRAV \citep{Richardson00,Stadel01} to model the gravitational interactions between planetesimals and giant planet cores. This version of PKDGRAV includes the detailed empirically derived analytical collision model (EDACM) from \citet{Leinhardt12} and \citet{Leinhardt15}. EDACM provides realistic outcomes for collisions in the gravity-dominated regime, although it neglects the effects of vaporization and the presence of the surrounding gas, which can change the outcomes of impacts \citep{StewartLPSC19a,Davies19}.

We examine five scenarios in which Jupiter and Saturn grew in-situ (`growth'), five in which the fully grown gas giants migrated inwards through the disk (`migration'), and six scenarios based on the Grand Tack model \citep{Walsh11}, in which Jupiter and Saturn grew and migrated inwards then outwards through the disk (`GT'). We also ran control simulations for two scenarios in which the giant planet cores remained in place and did not grow. The complete list of all 66 simulations included in this work is given in Appendix \ref{a:tables}.

\begin{figure}
    \centering
    \includegraphics[width=\columnwidth]{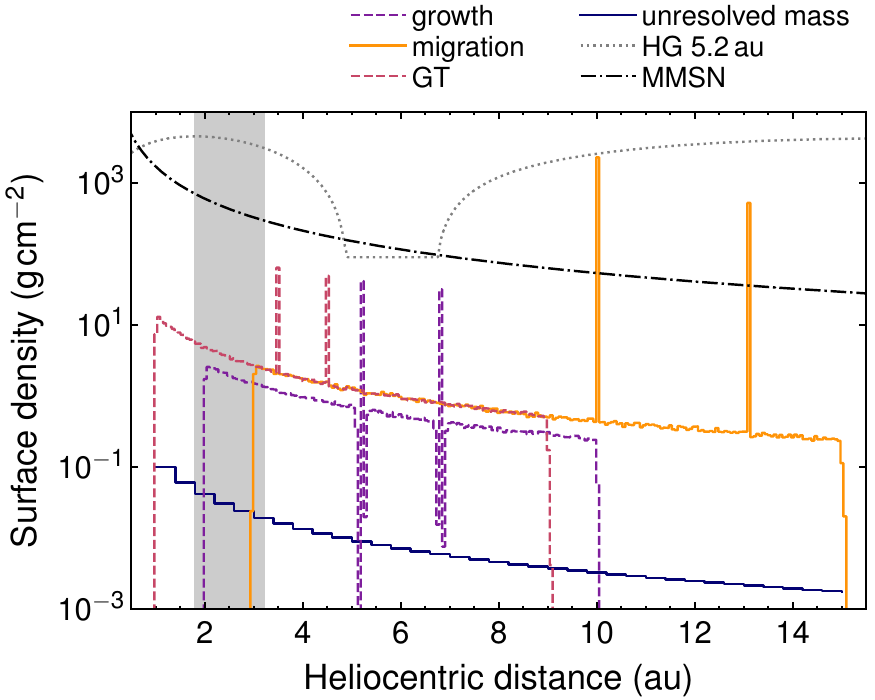}
    \caption{Surface density of solids (solid and dashed colored lines) and gas ({dotted and dash-dot} lines) used in the {various} simulations. The higher surface density gas disk based on hydrodynamic simulations (HG) is scaled according to the location of Jupiter, {such that the gap moves with Jupiter}. In this figure the HG profile is shown for Jupiter at 5.2\,au. {The initial unresolved mass was the same in all simulations.} The asteroid belt region (defined here as the region between 1.8 and 3.2\,au) is indicated by the grey band. The spikes in solid mass are due to the giant planets (cores).}
    \label{f:surfden}
\end{figure}

The simulations were initiated with either 20\,000, 50\,000, or 100\,000 equal mass planetesimals on cold orbits, and distributed based on the minimum mass solar nebula  (MMSN) model. The planetesimal surface density, $\Sigma_\mathrm{p}$, was defined according to,
\begin{equation}
    \Sigma_\mathrm{p} = \sigma_1 \left(\frac{r}{1\,\mathrm{au}}\right)^{-1.5}\,\mathrm{g\,cm^{-2}}, 
\end{equation}
where $\sigma_1$, the surface density at 1\,au, is 7\,g\,cm$^{-2}$ for the growth simulations, and 14\,g\,cm$^{-2}$ for the migration and Grand Tack simulations. 
Note that the mass of solid material incorporated into the giant planets was not removed from this planetesimal disk, effectively leaving an excess of mass in the vicinity of the giant planets. The growth cases included narrow gaps around the giant planet proto-cores. The simulations all began with a small mass of `unresolved' material distributed in annular bins (see below), representing smaller planetesimals and dust, following the same prescription as above with surface density at 1\,au of 0.1\,g\,cm$^{-2}$. In reality the fraction of solid mass already incorporated into planetesimals at this time may have been lower, and the corresponding mass remaining in dust or other small particles higher. The surface densities of solid materials for the three simulation types are shown as solid lines in Figure \ref{f:surfden}.

Planetesimals had densities of 2\,g\,cm$^{-3}$, appropriate for mixtures of metals, silicates and ices. The planetesimal disk was embedded in the potential well of a solar mass star. The sizes of planetesimals in our simulations were determined by the total disk mass and the number of particles. We obtain radii appropriate for large primitive planetesimals (350--600\,km). Aerodynamic drag depends on planetesimal size, and \citet{Raymond17} showed that the efficiency of implantation into the asteroid belt also depends on planetesimal size. Our planetesimals are large, but they are similar to the upper range of expected birth sizes \citep[e.g.][]{Simon16,Delbo17}, where the largest bodies should contain a large fraction of the total mass in planetesimals. We expect that planetesimals were forming in our solar system throughout the modeled time period, but we neglect ongoing planetesimal formation due to considerable uncertainty in the process and rates. 

Gas giant planets began either with their modern masses (migration simulations) or as cores of 3\,\mearth\ each (growth and GT simulations). Giant planets grew linearly in mass with accretion timescales of 20, 50, or 100\,kyr. {The real growth curve of a giant planet is expected to be more complex, but we keep the linear growth simplification used by previous works \citep{Walsh11,Raymond17}. This growth was imposed artificially, and the mass was not removed from the gas or planetesimal disks.} Migration was implemented as in \citet{Walsh11} and \citet{Carter15} by adding an acceleration to the giant planets to force them to achieve the Keplerian velocity corresponding to their desired location. We explored migration timescales of 50, 100, and 200\,kyr. Inward migration was linear in time, while the velocity change during outward migration in the Grand Tack scenarios followed an exponential function \citep[see][{note that the giant planets therefore do not reach their final locations in the duration of these simulations}]{Walsh11,Carter15}. In the growth simulations and migration simulations the planetesimal disk was allowed to begin evolving and to relax from the addition of the giant planets (cores) for a time, $t_\mathrm{start,J} =$ 5--50\,kyr, before the growth/migration began. The start times of these simulations, relative to the beginning of solar system history, is expected to be in the range 1--5\,Myr after CAIs.

\begin{table*}
    \centering
    \caption{Summary of simulation categories and their parameters. The start times of these simulations after the start of the solar system as measured by calcium-aluminum-rich inclusions (CAIs) are expected to be in the range 1--5\,Myr. $a_{\mathrm{J/S},0}$ give the initial location for Jupiter and Saturn (or their cores), $a_{\mathrm{J/S, min}}$ and $a_{\mathrm{J/S,end}}$ give the minimum and final heliocentric distances of the giant planets.
    \label{t:simcat}}
    \begin{tabular}{lrrrrrrrrrrr}
    \hline
        & Growth/migration  & Simulation    & Planetesimal & Inner & Outer & & & & & & \\
    Simulation type & timescales & durations & disk mass  &  edge   &  edge  & $a_{\mathrm{J},0}$	& $a_{\mathrm{S},0}$  & $a_{\mathrm{J, min}}$	& $a_{\mathrm{S, min}}$ & $a_{\mathrm{J,end}}$ & $a_{\mathrm{S,end}}$ \\
     & (kyr) & (kyr) & (\mearth)  & (au)   &  (au)  & (au)	& (au)  & (au)	&	(au) & (au) & (au) \\
    \hline
    growth        & 20, 50, 100 & 60, 130, 300 & 6   & 2   & 10  & 5.2	& 6.82  & --	& --	& --	& -- \\
    migration     & 50, 100, 200 & 150, 200, 300 & 14   & 3   & 15 & 10.0  & 13.1 & 5.2 & 6.82 & 5.2 & 6.82 \\
    GT 1.5\,au    & 50, 100 & 250, 500 & 13   & 1   & 9  & 3.5  & 4.5 & 1.5 & 1.97 & 5.2 & 6.82 \\
    GT 2.0\,au    & 50, 100 & 250, 500 & 13   & 1   & 9  & 3.5  & 4.5 & 2.0 & 2.62 & 5.2 & 6.82 \\
    \hline
    \end{tabular}
\end{table*}

All bodies (planetesimals and giant planets [cores]) gravitationally interacted with all other bodies, and experienced aerodynamic drag from the nebular gas based on \citet{Adachi76} and \citet{Brasser07} as described in \citet{Carter15}. We examine two different nebular gas profiles (see Figure \ref{f:surfden}), one based on the MMSN, the other accounting for gaps opened by the giant planets (labeled `HG'), based on hydrodynamical simulations from \citet{Morbi07}, as used in several previous works (e.g. \citealt{Walsh11,Raymond17}; see \citealt{Carter15} and Appendix \ref{a:gas} for details of the implementation). {The radial profile of the MMSN disk remained constant in simulations using this gas disk, the HG disk was scaled to the location of Jupiter such that the gap moves with the migrating giant planet.} The gas disk dissipated exponentially on a timescale, $\tau_\mathrm{gas}$, which varied from 50--200\,kyr. The orbits of bodies were also subject to tidal damping (but not migration) due to interactions with the nebular gas based on the prescription, and the inclination and eccentricity damping timescales, given by \citet{Cresswell08}. 

We expect a distribution of planetesimal and particle sizes for collision fragments \citep[see][]{Leinhardt12}. As EDACM can produce many fragments as a result of a collision, it is necessary, to maintain computational practicality, to impose a mass limit, $M_\mathrm{min}$, below which planetesimals are no longer resolved. Collision remnants smaller than this minimum mass are treated semi-analytically as `unresolved debris', distributed in 0.4\,au wide annular bins. This simple treatment of small debris ignores mutual collisions amongst the debris, formation of new planetesimals, and drag processes that may act on the small bodies or particles this debris represents. The fully resolved bodies accrete the unresolved mass as they pass through each bin \citep[see][for more details]{Leinhardt05,Leinhardt15,Carter15}, `recycling' this material. Some of our growth (and control) simulations had this minimum mass set too high (greater than the initial planetesimal mass), resulting in most collisions producing no remnants and artificially increasing the debris mass. The collision numbers in the growth simulations (both with an without this problem) are sufficiently low that very few bodies would have collided multiple times, and this resolution limit therefore has a negligible effect on the overall results (the simulations with high remnant mass limits were, however, excluded from the collision outcome frequencies given in Table \ref{t:summary}).

In some previous works \citep[e.g.][]{Kokubo+Ida02,Carter15}, planetesimal radius expansion has been used to decrease the collision and evolution timescales. In this work, we chose not to apply radius expansion as the durations of the simulations are sufficiently short for reasonable computation times. 

The disk mass, extent, and decay timescale, the starting locations of Jupiter and Saturn, the planetesimal minimum mass, and the simulation timestep and duration were all chosen based on the specific scenario for computational efficiency. For a simulation with a fixed timestep there is a minimum heliocentric distance inside which orbits will not be well resolved, we ignore collisions that occurred closer to the Sun than 1.5, 1.5, and 0.7\,au for the growth, migration and Grand Tack simulations respectively. Full details of the simulation parameters are given in Tables \ref{t:simcat}, \ref{t:sims}, \ref{t:sims2}, and \ref{t:sims3}.

%%%%%%%%%%%%%%%%%%%% RESULTS %%%%%%%%%%%%%%%%%%%%%

\section{Results}

\subsection{Planetesimal disk evolution}

\begin{figure}
%\begin{interactive}{animation}{F2_growthis1e5_100k_hg1_aeiv_mov.mp4}
\includegraphics[width=\columnwidth]{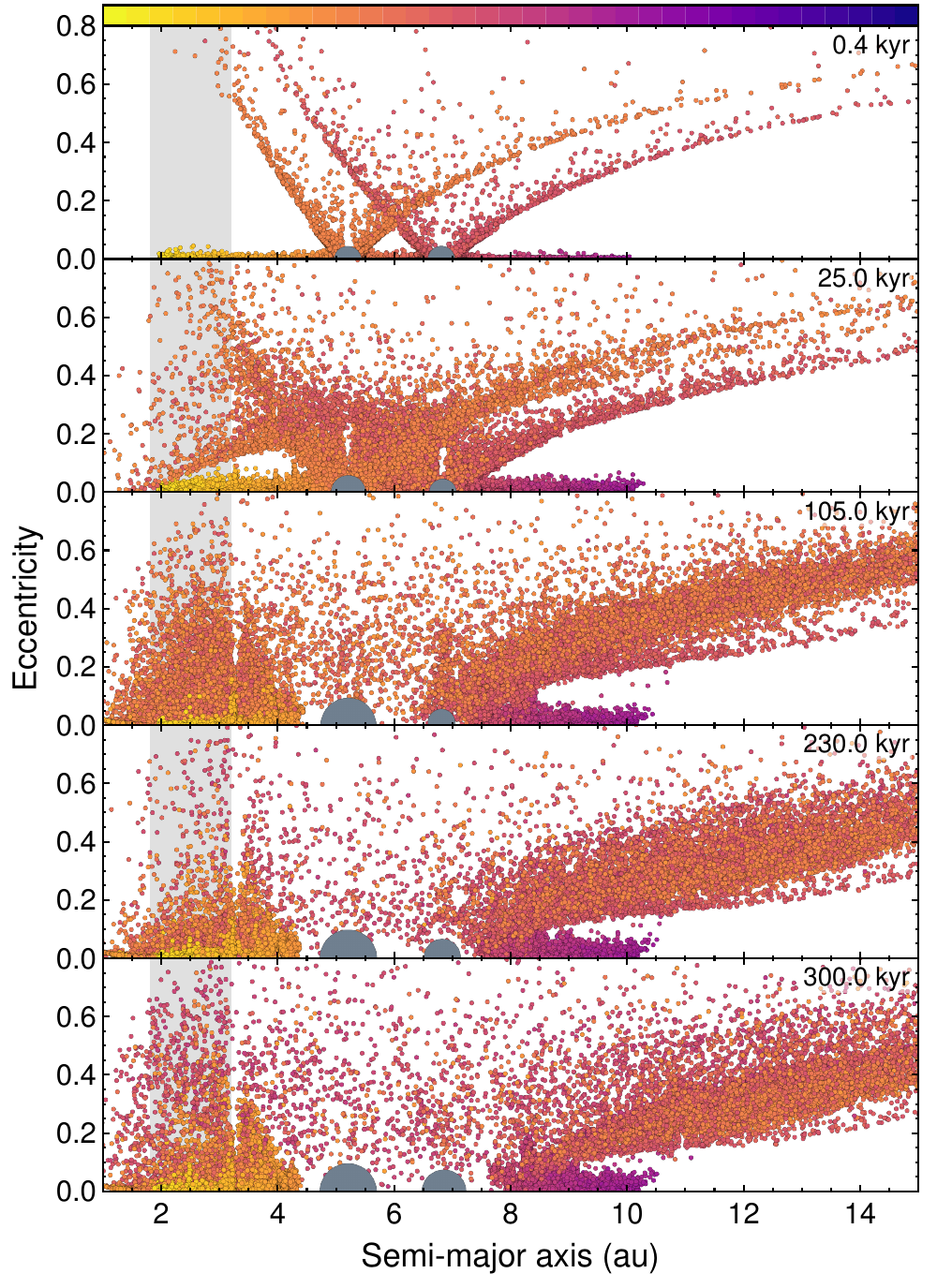}
%\end{interactive}
    \caption{Planetesimal disk eccentricity evolution through five time snapshots in an example in-situ growth simulation with a 100\,kyr growth timescale and HG gas disk (growthis1e5\_100k\_hg1 {in Table \ref{t:sims}}). Planetesimals are colored according to their average composition as a function of initial heliocentric distance of constituent material; (proto-)Jupiter and (proto-)Saturn are shown as grey circles. Body sizes are proportional to their mass. The light grey band indicates the present-day asteroid belt region. 
    An animation of this figure is available, it shows eccentricities and inclinations of the bodies and impact velocities vs semi-major axis for the 300\,kyr duration of the simulation. The duration of the video is 38 seconds.}
    \label{f:growthae}
\end{figure}

As expected, giant planet growth and migration cause substantial excitation of planetesimal orbits in our simulations. The evolution of a planetesimal disk in the presence of Jupiter and Saturn as they grow from 3\,\mearth\ cores into gas giants is shown in Figure \ref{f:growthae}. At the beginning of the simulation the static cores excite nearby planetesimals along `wings' in eccentricity--semi-major axis space corresponding to constant Tisserand parameter \citep{Raymond17}. Most planetesimals exhibit modest excitations of up to an eccentricity of $\sim$0.1 in this early stage, as seen by \citet{Raymond17}. However, we observe a population of planetesimals that have substantially higher eccentricities in these early wings. We expect that the higher eccentricity planetesimals are caused by planetesimal self-excitation: Unlike in previous studies, in our simulations the planetesimals gravitationally (and collisionally) interact with each other allowing further excitation not observed in other works.

These extended wings rapidly `droop' over time, with the planetesimal eccentricities falling due to aerodynamic drag from the nebular gas. As Jupiter begins growing rapidly, at 20\,kyr in this case, a second wave of greater planetesimal excitation begins. Jupiter's growth leads to the `double wing' structure seen around 3-4\,au in the second panel of Figure \ref{f:growthae}. Jupiter's continued growth and the later growth of Saturn cause further orbit excitation and scattering of planetesimals. Gas drag acts to circularize the orbits of the planetesimals, causing many of those that began near Jupiter to end up in the region of the asteroid belt, or on orbits that cross the region in which the terrestrial planets would be growing. This evolution is similar to that seen by \citeauthor{Raymond17} (see their figure 2), though, again, we see some planetesimals with much higher eccentricities.

The general phenomenology seen in Figure \ref{f:growthae} is common across all the growth simulations, however, the emergence of the `double wing' structure is dependent on the nebular gas profile and the planetesimal size used. Weaker drag experienced by larger planetesimals, faster growth timescales, or planetesimals in lower density gas substantially reduce the appearance and deflection of the lower {eccentricity} wing. As such, the lower wing is only visible in the higher resolution, slower growth simulations with a hydrodynamic gas profile. Further examples of the evolution during planet growth are provided in appendix \ref{a:examples}.

We also examined control cases in which the giant planet cores remained at their initial locations and masses (3\,\mearth). In these simulations, the initial excitation of the disk is very similar to that seen in the first panel of Figure \ref{f:growthae}. However, the excitation does not grow significantly larger, and the planetesimal orbits are damped over time by drag, resulting in much lower excitation for planetesimals that have semi-major axes in the asteroid belt region at the end of the simulation. An example of the evolution in a control simulation is provided in appendix \ref{a:control}.

\begin{figure}
%\begin{interactive}{animation}{F3_mig10in1e5_100k_hg1_aeiv_mov.mp4}
\includegraphics[width=\columnwidth]{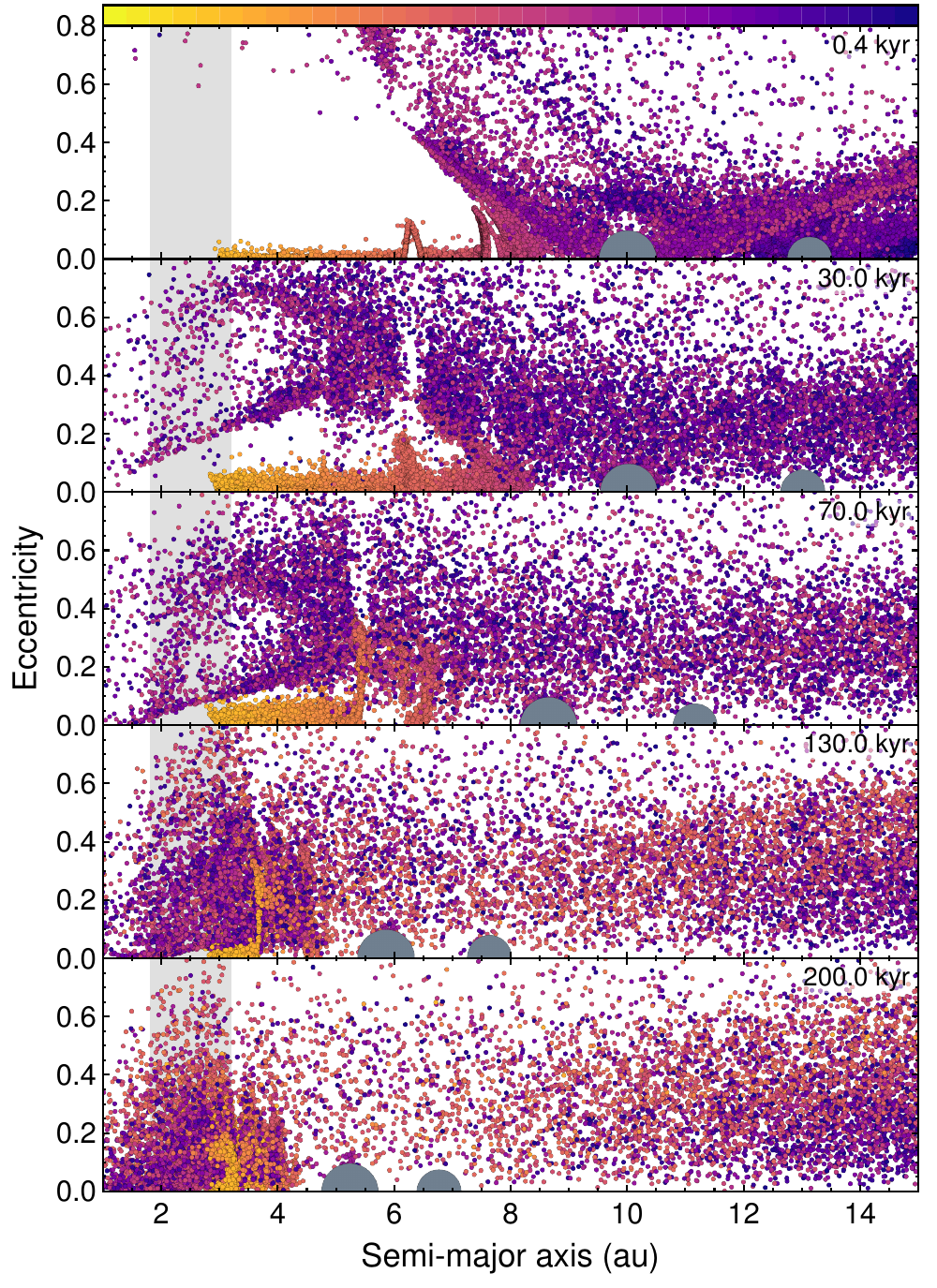}
%\end{interactive}
    \caption{Planetesimal disk eccentricity evolution through five time snapshots in an example migration simulation with a 100\,kyr migration timescale and HG gas disk (mig10in1e5\_100k\_hg1 {in Table \ref{t:sims}}). Planetesimals are colored according to their average composition as a function of initial heliocentric distance of constituent material; Jupiter and Saturn are shown as grey circles.  The light grey band indicates the present-day asteroid belt region. 
    An animation of this figure is available, it shows eccentricities and inclinations of the bodies and impact velocities vs semi-major axis for the 200\,kyr duration of the simulation. The duration of the video is 25 seconds.}
    \label{f:migae}
\end{figure}
\begin{figure}
%\begin{interactive}{animation}{F4_GT15gm5e4_50k_hg1_aeiv_mov.mp4}
\includegraphics[width=\columnwidth]{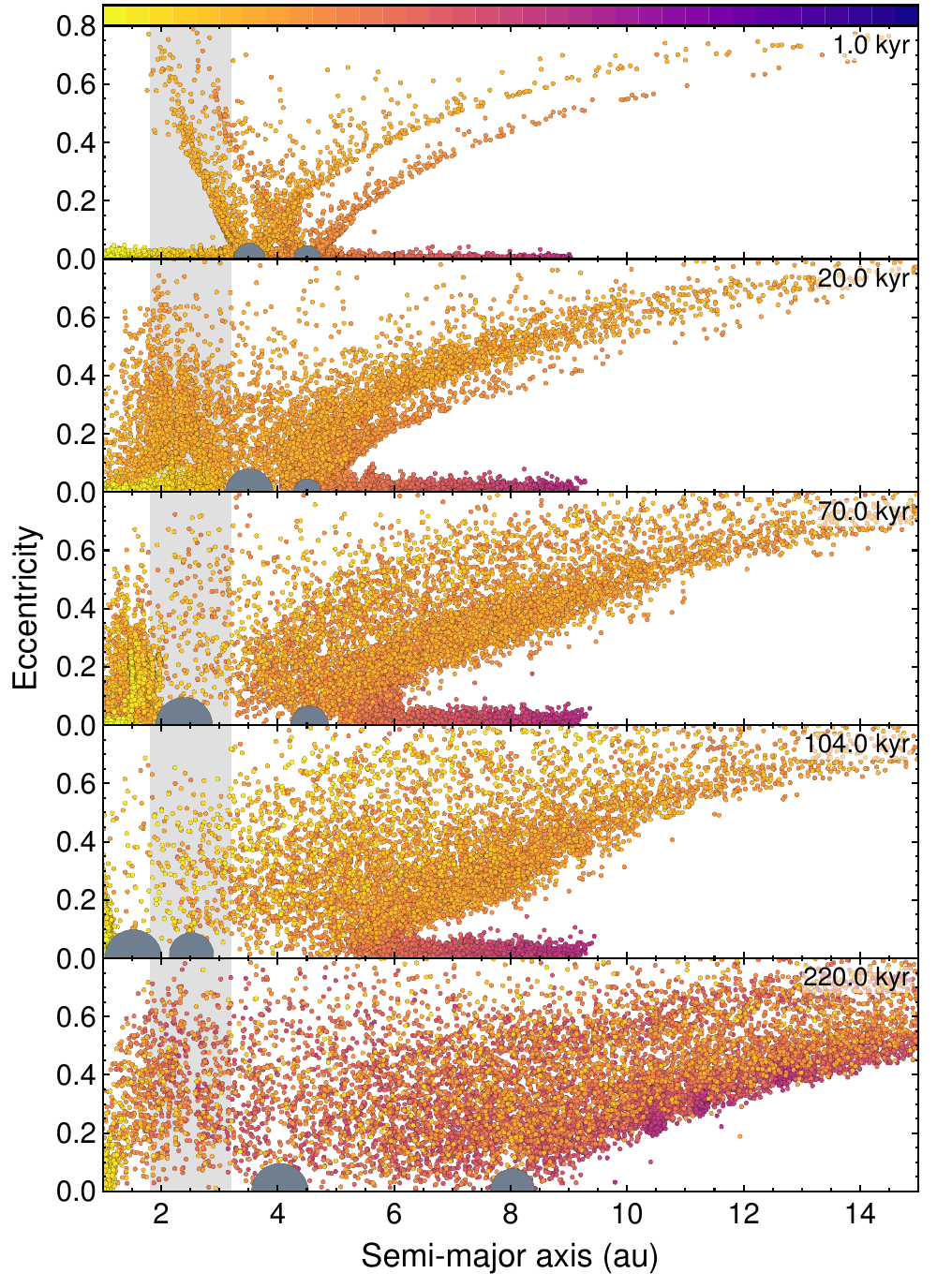}
%\end{interactive}
    \caption{Planetesimal disk eccentricity evolution through five time snapshots in an example Grand Tack simulation with a 50\,kyr evolution timescale and HG gas disk (GT15gm5e4\_50k\_hg1 {in Table \ref{t:sims}}). Planetesimals are colored according to their average composition as a function of initial heliocentric distance of constituent material; (proto-)Jupiter and (proto-)Saturn are shown as grey circles. Body sizes are proportional to their mass.  The light grey band indicates the present-day asteroid belt region. 
    An animation of this figure is available, it shows eccentricities and inclinations of the bodies and impact velocities vs semi-major axis for the 250\,kyr duration of the simulation. The duration of the video is 50 seconds.}
    \label{f:GTae}
\end{figure}

In our `migration only' simulations, Jupiter and Saturn begin at 10.0 and 13.1\,au with their final masses, and migrate inwards. An example of the planetesimal disk evolution in this scenario, in terms of planetesimal eccentricity as a function of semi-major axis, is shown in Figure \ref{f:migae}. The much more massive initial planets cause substantially larger wings with higher eccentricities, and large mean motion resonance features (visible at $\sim$6.3 and $\sim$7.6\,au in the first panel of Figure \ref{f:migae}). The sudden introduction of massive giant planets into the planetesimal disk may cause some unrealistic extra excitation \citep[see][]{Dobinson16}; when we consider impact velocities below, we ignore impacts that occurred in the initial 50\,kyr `burn-in' period before the migration began.

The double wing structure observed in the growth simulations is seen again in our migration simulations as the planets begin migrating inward, and the resonance features broaden. The planetesimals captured in these resonances are then shepherded inwards in walls as Jupiter continues to move towards the inner disk (visible at $\sim$5.4 and $\sim$6.5\,au in the third panel of Figure \ref{f:migae}; \citealt{Tanaka99,Fogg05,Mandell07}). While the eccentricities of planetesimals are substantially enhanced by these strong mean motion resonances, the inclinations are (in general) not (see section \ref{s:orbexcite}). These eccentricity walls lead to increased collision rates and impact velocities, as well as efficiently moving material closer to the central star. The eccentricities in the inner disk at late times resemble those for the in-situ growth scenario shown in Figure \ref{f:growthae}, though with some clear bands due to resonances with the giant planets. The inner disk also contains a much greater mass of material that originated from more distant parts of the protoplanetary disk (beyond $\sim$8\,au, purple and blue points in Figures \ref{f:growthae} and \ref{f:migae}).

The migration scenarios we simulated show similar results to those found by \citet{Raymond17}, though it appears that we see more outer disk material (from beyond Jupiter's initial orbit) scattered into the inner disk. This may be due to our choice to begin these simulations with fully grown giant planets, rather than to grow them from cores first. 
These migration only simulations result in wet, outer solar system planetesimals (blue and purple points in Figure \ref{f:migae}) being implanted throughout the asteroid belt. 
Note that the lack of planetesimals interior to 3\,au in the initial setup of these simulations means we cannot interpret the structure of the asteroid belt for proportions of wet vs. dry, inner solar system planetesimals.

An example of planetesimal disk evolution in the Grand Tack scenario is shown in Figure \ref{f:GTae}. With growth followed by migration, the Grand Tack simulations show similarities to both the growth and migration scenarios (see Figures \ref{f:growthae} and \ref{f:migae}). Before migration begins there is significant scattering into the inner disk due to the growth of the gas giants, this is followed by further scattering and shepherding of walls of planetesimals towards the Sun. The final outward migration phase scatters more planetesimals, in particular delivering planetesimals from beyond Jupiter's and Saturn's initial orbits into the inner disk \citep[e.g.][]{Walsh11}, as well as pushing some outer solar system planetesimals further from the Sun in resonant walls. Further examples of the evolution of planetesimal disks during planet growth and migration are provided in appendix \ref{a:examples}.

\subsection{Planetesimal impact velocities}

\begin{figure}
\includegraphics[width=\columnwidth]{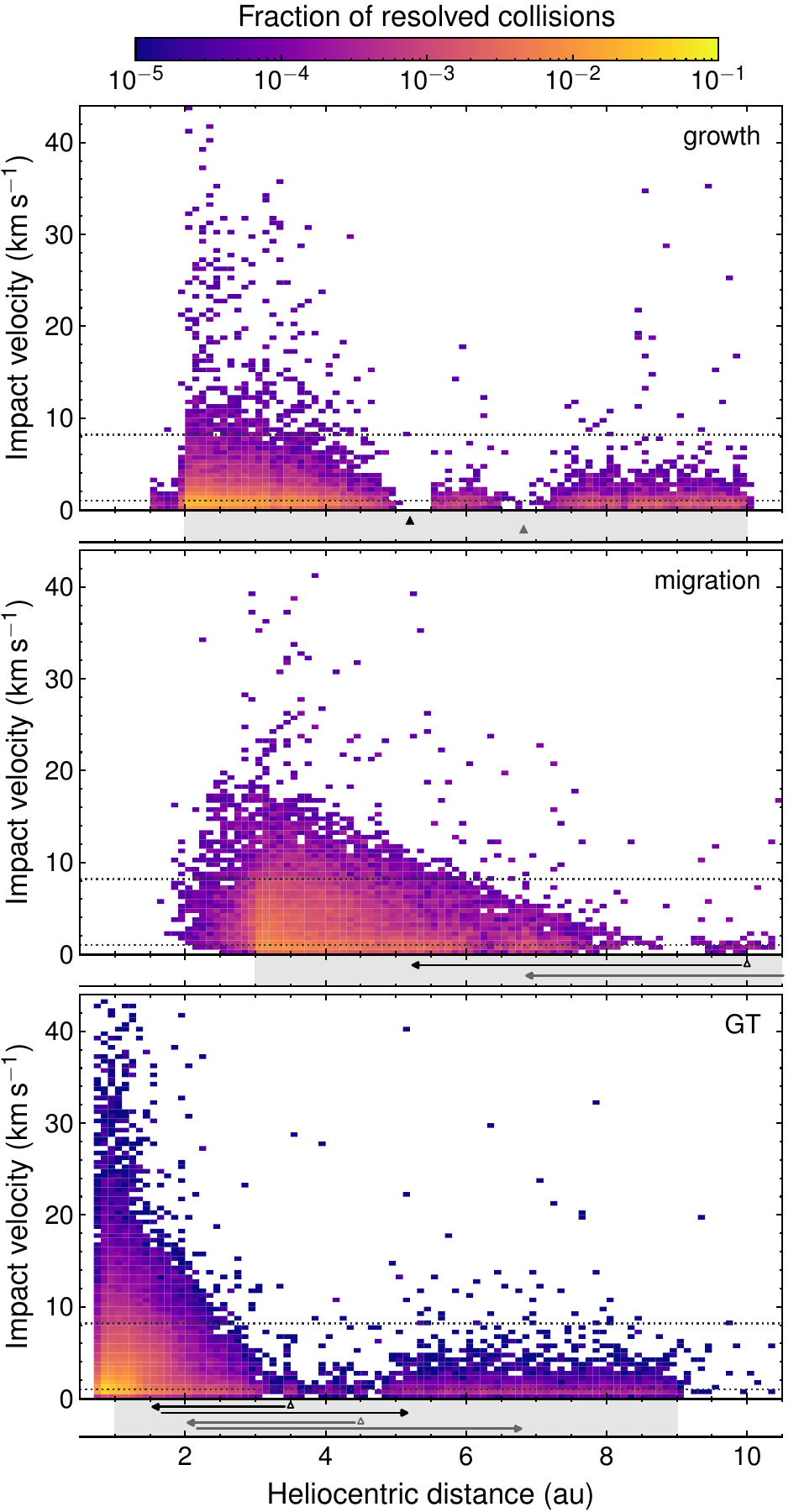}
    \caption{Impact velocity distributions as a function of heliocentric distance for the three simulation groups. All simulations for each simulation type are combined in this figure. Blue colors indicate few collisions in a bin, orange colors indicate the locations of the majority of collisions in the simulations. The dashed lines indicate impact vaporization thresholds for ice ($\sim$1\,km\,s$^{-1}$) and warm forsterite ($\sim$8\,km\,s$^{-1}$). The area below each panel indicates the location and migration of Jupiter (black) and Saturn (grey). Solid triangles indicate the fixed positions of the giant planets, open triangles indicate initial positions from which the giant planets followed the path indicated by the arrows. The light grey bands indicate the initial bounds of the planetesimal disks.}
    \label{f:avhist}
\end{figure}

We now turn our attention to planetesimal-planetesimal collisions. Histograms of planetesimal impact velocities for the three types of simulations are shown in Figure \ref{f:avhist}. The impact velocities and collision statistics given throughout this work do not include impacts between planetesimals and giant planets (cores), which are assumed to be accretionary. The growth and migration scenarios both include a `burn-in' period before growth/migration begins (prior to $t_\mathrm{start,J}$), we ignore these periods in our analysis of impact velocities in order to minimize effects not associated with the growth/migration itself.

A larger fraction of all the resolved planetesimal collisions that occur in any simulation occur closer to the Sun where orbital timescales are shorter and disk surface densities are higher. This increased collision rate closer to the central star accounts for much of the structure in the growth simulations' velocity distribution. Planetesimals orbiting at smaller heliocentric distances will also have larger random velocities, which are proportional to the higher Keplerian orbital velocities, leading to higher impact velocities closer to the star.

\begin{figure*}
\includegraphics[width=\textwidth]{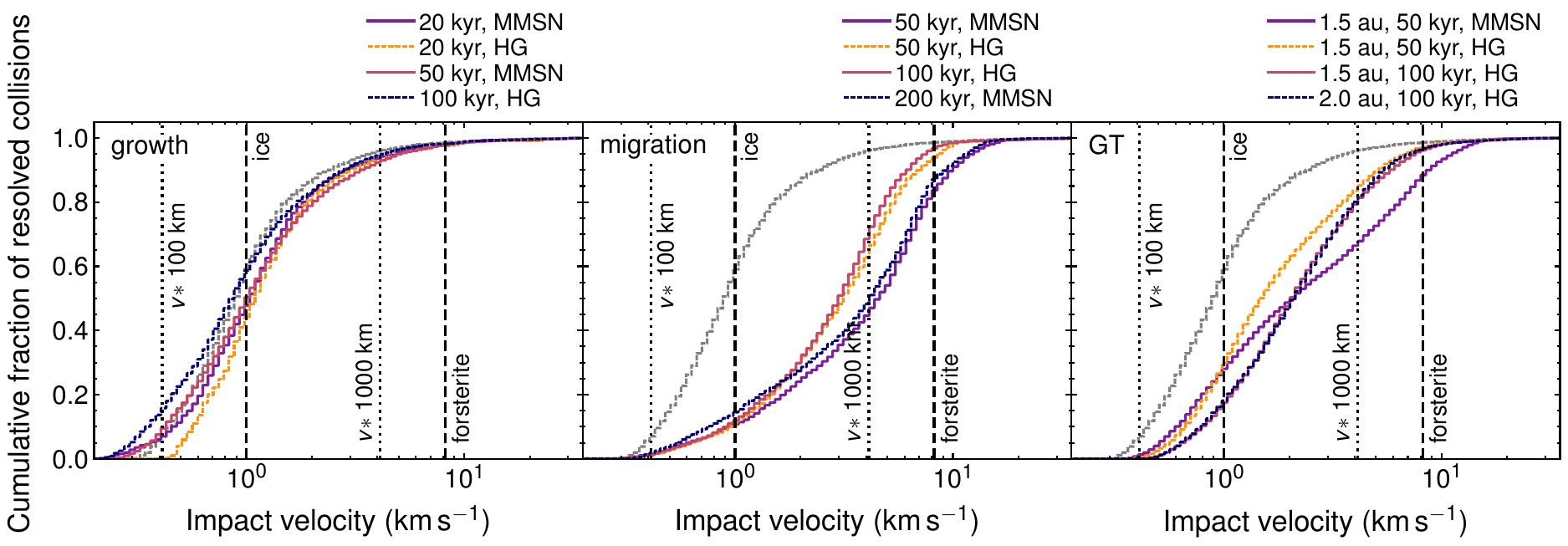}
    \caption{Cumulative impact velocity distributions after the start of growth or migration for the three simulation groups. The grey dashed histogram shows the cumulative impact velocity distribution for the control {simulations (without giant planet growth)}. The dashed vertical lines indicate impact vaporization thresholds for ice ($\sim$1\,km\,s$^{-1}$) and warm forsterite ($\sim$8\,km\,s$^{-1}$), the dotted lines indicate catastrophic disruption thresholds for head-on equal mass impacts between planetesimals with radii of 100\,km and 1000\,km (based on \citealt{Leinhardt12}). The variation at the low velocity end of the distributions is due to differences in planetesimal mass between different resolution simulations.}
    \label{f:velhistcumu}
\end{figure*}

There are obvious differences in the spatial distribution of collisions between the three simulation types shown in Figure \ref{f:avhist}, in part due to the different regions of the disk modeled in each case. The most striking feature is the large increase in high impact velocity collisions interior to the migrating Jupiter in the migration and Grand Tack simulations. The walls of excited planetesimals cause substantially more, and higher impact velocity, collisions close to the locations of these walls. In the case of the Grand Tack, Jupiter's shepherding results in a huge spike in impact velocities and numbers of collisions at $\sim$1\,au. This spike in the Grand Tack case is established by design, the tack point was chosen to mold the inner disk in to a narrow annulus where most of the mass ends up concentrated inside the orbit of Mars \citep{Walsh11}. Also very noticeable is the dearth of collisions close to the initial locations of the giant planets (cores). This lack of collisions is not surprising, most of the planetesimals close to Jupiter and Saturn are rapidly scattered away from these locations or accreted by the giant planets.

Cumulative velocity distributions for several simulation groups in all three categories are shown in Figure \ref{f:velhistcumu} (indicated by color). Also shown, in grey, is the cumulative velocity distribution for the control simulations in which no growth or migration occurred. This grey line is largely obscured in the left hand panel of Figure \ref{f:velhistcumu}, revealing that the growth of giant planets has little effect on the impact velocities throughout the disk. The variation at the low velocity end of these distributions, which is particularly noticeable in the growth case, is due to differences in the initial planetesimal mass between simulations with different resolutions (numbers of particles). The minimum values in the impact velocity distributions are close to the mutual escape velocity for two of the least massive planetesimals, which is lower for the lower mass bodies found in higher resolution simulations.

In all cases there are a significant number of collisions ($>$40\%) that exceed a velocity of 1\,km\,s$^{-1}$, the velocity threshold at which shock-vaporization of water ice begins \citep{Stewart08,Davies19b}. This criterion is the velocity required for the onset of vaporization in a 1D, symmetric impact upon release to pressures appropriate for the solar nebula. Collisions that exceed this impact velocity lead to partial vaporization of the shocked material.

The migration and Grand Tack simulations show substantially increased impact velocities compared to the growth and control models (see center and right hand panels of Figure \ref{f:velhistcumu}). In the growth simulations $\gtrsim$40\% of impacts exceed the threshold for vaporization of water ice, this rises to $\gtrsim$75\% for the migration and Grand Tack models. The tail of the impact velocity distribution also exceeds the vaporization threshold for warm rocky planetesimals ($\sim$8\,km\,s$^{-1}$ for forsterite, \citealt{Davies20}) in all cases ({$\sim$2\% of impacts for the growth simulations;} Figure \ref{f:avhist}), though there are clearly much larger proportions of these very high velocity collisions in the simulations that include gas giant migration {(3--18\%)}. The walls of high eccentricity planetesimals pushed inwards by the migrating giant planets are the key factor in increasing impact velocities. Planetesimals in these walls encounter the near-circular orbiting planetesimals ahead of them, leading to high relative velocities, especially closer to the star, where the Keplerian orbital velocities are greater.

\subsection{Collision outcomes}

\begin{sidewaystable*}
    \vspace{80mm}
    \centering
    \caption{Summary of global results for all simulation categories. {The critical velocities for vaporization of water ice and warm forsterite are 1\,km\,s$^{-1}$ and 8.2\,km\,s$^{-1}$ respectively \citep{Davies20}.} Collision outcomes listed are perfect mergers (merge), {partial accretion collisions (accrete),} erosive impacts (erode), and hit-and-run impacts (h-\&-r).
    \label{t:summary}}
    \begin{tabular}{lrcccccrrcccc}
    \hline
                    &     Time &  & & Standard & \multicolumn{2}{c}{Fraction of collisions} & \multicolumn{4}{c}{} & \multicolumn{2}{c}{Fraction of resolved} \\
                    &       scale          & Median impact      & Mean impact & deviation & \multicolumn{2}{c}{exceeding critical velocity for:} & \multicolumn{4}{c}{Resolved collision outcomes (\%)} & \multicolumn{2}{c}{planetesimals that:} \\
    Simulation type & (kyr)  & velocity (km\,s$^{-1}$) & velocity (km\,s$^{-1}$) & (km\,s$^{-1}$) & ice & forsterite   &  merge    & accrete & erode  & h-\&-r   &  collided     & survived \\
    \hline
    growth, MMSN        & 20   & 1.03 & 1.71 & 2.89  & 0.54 & 0.02   & 17 & 27 & 13 & 43 & 0.016  & 0.67 \\
    growth, HG          & 20   & 1.09 & 1.86 & 3.11  & 0.58 & 0.03   & 21 & 25 & 9 & 45 & 0.016  & 0.69 \\
    growth, MMSN        & 50   & 0.99 & 1.76 & 2.74  & 0.52 & 0.02   & 17 & 26 & 14 & 43 & 0.024  & 0.64 \\
    growth, MMSN        & 100  & 1.02 & 1.92 & 3.22  & 0.54 & 0.03   & 17 & 25 & 15 & 43 & 0.037  & 0.60 \\
    growth, HG          & 100  & 0.86 & 1.53 & 2.80  & 0.43 & 0.02   & 15 & 25 & 17 & 42 & 0.059  & 0.66 \\
    \hline
    no growth, MMSN     & --   & 0.90 & 1.40 & 2.16  & 0.43 & 0.01   & 19 & 25 & 7 & 49 & 0.028  &  0.96  \\
    no growth, HG       & --   & 0.93 & 1.43 & 2.17  & 0.44 & 0.02   & 25 & 29 & 9 & 37 & 0.028  &  0.96  \\
    \hline
    migration, MMSN     & 50   & 2.59 & 5.25 & 4.03  & 0.90 & 0.18   & 9 & 15 & 33 & 42 & 0.021  & 0.44 \\
    migration, HG       & 50   & 2.47 & 3.88 & 3.98  & 0.90 & 0.07   & 7 & 16 & 31 & 45 & 0.031  & 0.34 \\\
    migration, MMSN     & 100  & 2.58 & 5.10 & 6.05  & 0.87 & 0.17   & 9 & 15 & 34 & 42 & 0.025  & 0.42 \\
    migration, HG       & 100  & 2.52 & 3.60 & 3.85  & 0.89 & 0.04   & 7 & 15 & 30 & 47 & 0.034  & 0.36 \\
    migration, MMSN     & 200  & 3.06 & 4.83 & 3.83  & 0.86 & 0.14   & 8 & 14 & 38 & 40 & 0.031  & 0.38 \\
    \hline
    GT 1.5\,au, MMSN    & 50   & 2.12 & 3.74 & 3.79  & 0.73 & 0.12   & 8 & 16 & 31 & 45 & 0.144  & 0.09 \\
    GT 1.5\,au, HG      & 50   & 1.53 & 2.49 & 2.86  & 0.72 & 0.03   & 9 & 22 & 19 & 50 & 0.285  & 0.14 \\
    GT 2.0\,au, HG      & 50   & 1.63 & 2.52 & 2.90  & 0.78 & 0.03   & 8 & 24 & 17 & 51 & 0.233  & 0.31 \\
    GT 1.5\,au, MMSN    & 100  & 2.42 & 4.05 & 4.11  & 0.76 & 0.15   & 8 & 17 & 29 & 46 & 0.138  & 0.07 \\
    GT 1.5\,au, HG      & 100  & 2.11 & 2.97 & 2.93  & 0.84 & 0.04   & 6 & 21 & 20 & 53 & 0.274  & 0.09 \\
    GT 2.0\,au, HG      & 100  & 2.15 & 2.92 & 2.87  & 0.83 & 0.03   & 7 & 20 & 20 & 53 & 0.276  & 0.24 \\
    \hline
    \end{tabular}
\end{sidewaystable*}

The impact outcomes for each simulation category are listed in Table \ref{t:summary}. We list the percentages of impacts that lead to perfect merging, {partial accretion of the projectile, erosion of the target, and hit-and-run impacts.} The trends in the number of disruptive impacts match with the impact velocity distributions, migration-only simulations show the highest fractions of disruptive impacts, and Grand Tack simulations show more than growth simulations. It is clear that not all impacts disrupt both planetesimals in any scenario; {a large fraction of collisions were accretive} in all the simulations. 

The majority of impacts in all the simulations exceed the catastrophic disruption velocity ($v^*$) for head-on impacts between two $\sim$100\,km-size bodies. Catastrophic disruption occurs when the largest post-collision remnant contains less than 50\% of the total colliding mass \citep{Leinhardt12}. The frequency of collisions above this impact velocity suggests widespread disruption of colliding planetesimals. The dotted lines in Figure \ref{f:velhistcumu} indicate the catastrophic disruption velocity for head-on impacts, collisions at greater impact angles will be less efficient at disruption \citep{Leinhardt12}. There are still some merging impacts, and hence growth, even in the most excited scenarios. Note that our choice to initialize the simulations with equal size planetesimals likely substantially increased the number of hit-and-run impacts compared to an initial power law size distribution.

While some collisions result in the conversion of most of the impacting masses into unresolved debris, due to reaccretion, the unresolved debris is never a substantial fraction of the total planetesimal mass. Though it should be noted that this debris reaccretion efficiency may be too high (see section \ref{s:chondruleDiscuss}). The cumulative mass ejected as unresolved fragments is $\sim$1\% in the growth and migration simulations, but can reach as high as 10--15\% in the Grand Tack simulations. The mass ejected as unresolved debris is highest in the inner parts of the disk where most of the collisions occur.

\subsection{The fates of planetesimals}

\begin{figure}
\includegraphics[width=\columnwidth]{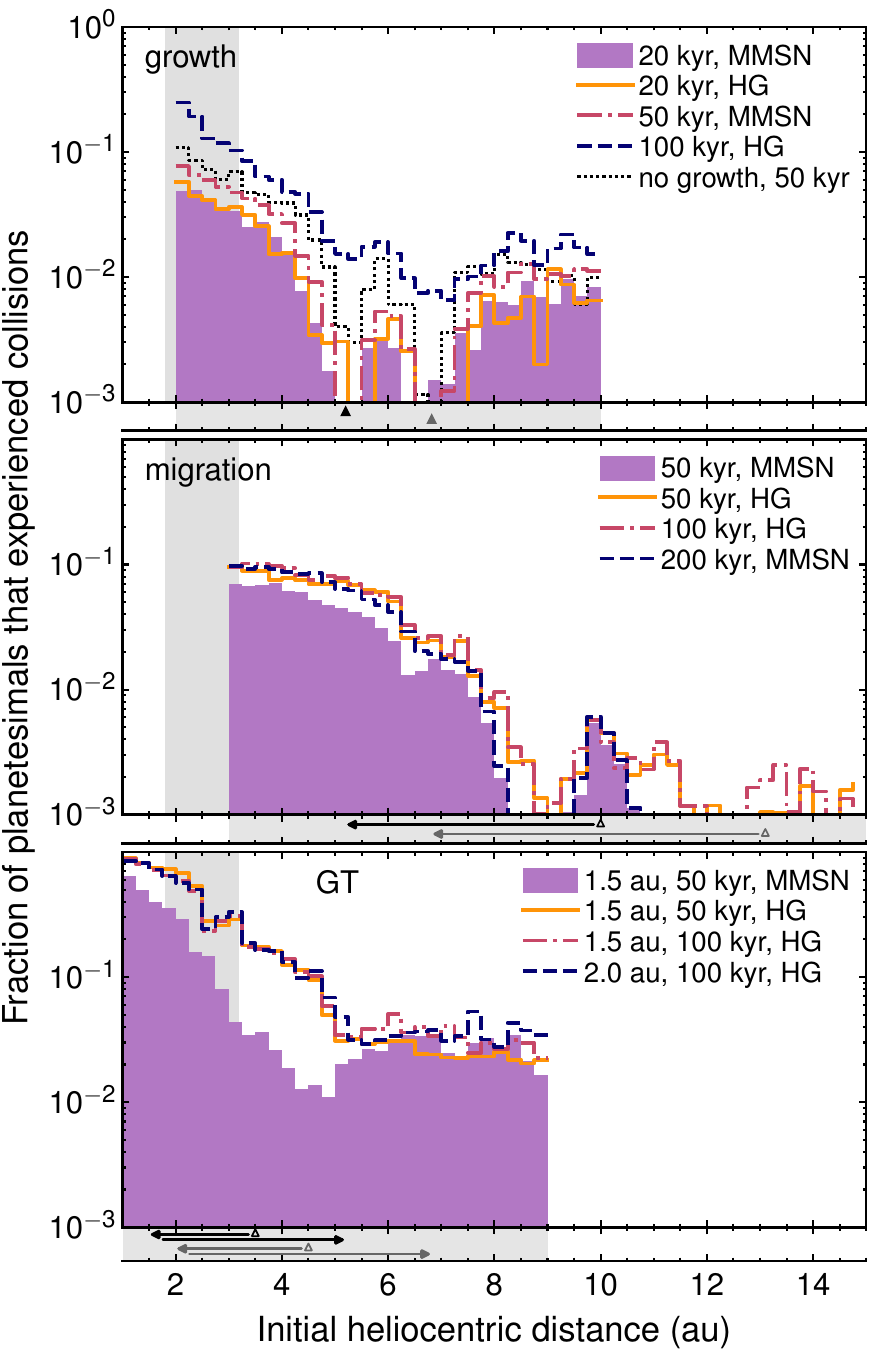}
    \caption{Fraction of planetesimals that experienced collisions as a function of their initial location within the disk. These collisions could have occurred anywhere in the disk. The collision fractions are averaged across similar simulations. The top panel shows the results from in-situ growth simulations, the middle panel migration-only simulations, and the bottom panel Grand Tack simulations. The area below each panel indicates the location and migration of Jupiter (black) and Saturn (grey). Solid triangles indicate the fixed positions of the giant planets, open triangles indicate initial positions from which the giant planets followed the path indicated by the arrows. The light grey bands below the axes indicate the bounds of the initial planetesimal disks. Note that the solid histogram is used for easier comparison and generally shows the lowest values, it should not be considered the nominal case.}
    \label{f:FracColl}
\end{figure}
\begin{figure}
\includegraphics[width=\columnwidth]{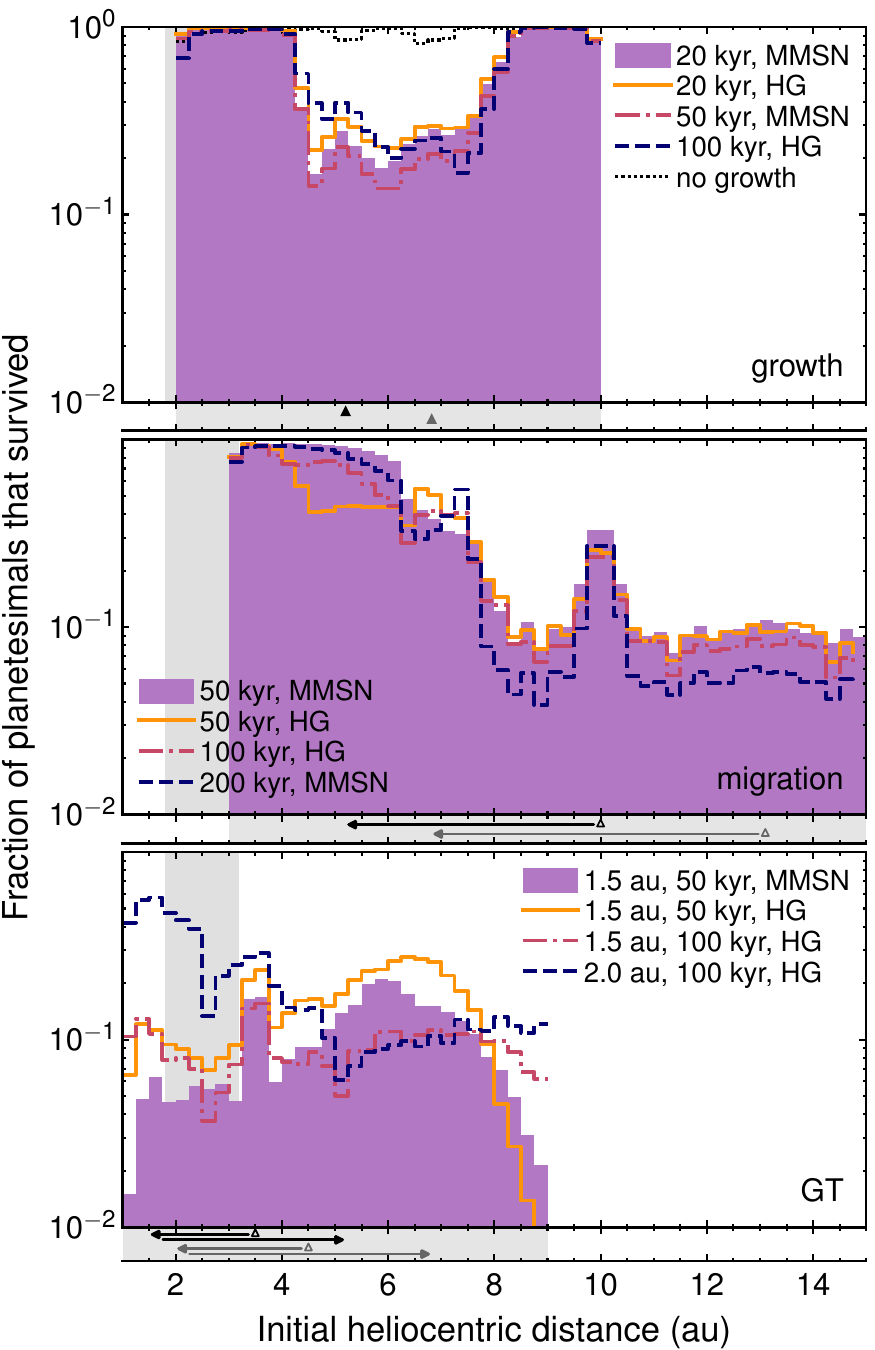}
    \caption{Fraction of planetesimals that survived to the end of the simulations as a function of their initial location within the disk. We consider planetesimals to have survived only if they remained within the radii of the modeled disk as shown by the grey bands below the $x$-axes. Planetesimals that began close to, or in the path of, the gas giants and survived were mostly scattered or shepherded to other regions of the disk. The survival fractions are averaged across similar simulations. The top panel shows the results from in-situ growth simulations, the middle panel migration-only simulations, and the bottom panel Grand Tack simulations. The area below each panel indicates the initial location and migration of Jupiter (black) and Saturn (grey) as in Figure \ref{f:FracColl}. Note that the solid histogram is used for easier comparison and generally shows the lowest values, it should not be considered the nominal case.}
    \label{f:FracSurvive}
\end{figure}

We have seen that planetesimals can reach high eccentricities due to interactions with growing and migrating giant planets, and that velocities of collisions between planetesimals can be extremely large. Two key questions are how many planetesimals collide, and how many survive these periods of giant planet growth and migration? \citet{Batygin15} suggested that the collisions induced by the Grand Tack would destroy the majority of planetesimals interior to Jupiter's initial orbit (causing their mass to be lost from the disk via aerodynamic drag into the Sun), while many authors have neglected the effects of planetesimal-planetesimal collisions entirely. In this section, we examine how many resolved planetesimals undergo collisions, and how many survive the period of giant planet growth and migration.

Figure \ref{f:FracColl} shows the proportion of resolved planetesimals that underwent at least one collision during the simulations as a function of their initial location within the disk. There is some variation with growth and migration timescales, and gas disk, but the major difference in the fraction of planetesimals that collided is between the different giant planet evolution scenarios (see also Table \ref{t:summary}). Giant planet migration induces more collisions than growth alone.

The reader should recall that the extent of the simulated disks varied between the simulations, and that this complicates the comparisons between the different models when considering the global averages given in Table \ref{t:summary}. The choice of disk extent was made to keep the simulations as computationally efficient as possible. The comparisons between corresponding bins in planetesimal heliocentric distance shown in Figure \ref{f:FracColl} (and Figures \ref{f:FracSurvive} and \ref{f:FracImplant}) give a clearer view of planetesimal fates.

Longer growth timescales are modeled with longer duration simulations, and therefore a greater length of time in which collisions can occur, thus longer growth timescales show somewhat higher fractions of resolved planetesimals that experience collisions. In all simulations there is a clear increase in collision fraction closer to the central star. Closer to the star the planetesimal surface density is greater, and so planetesimals are more likely to collide in any given time interval.

In the in-situ growth simulations, the fraction of resolved planetesimals experiencing collisions reaches its maximum of $\sim$10\% near the inner edge of the modeled disk ($\sim$3\,au; Figure \ref{f:FracColl}), and only in cases with longer growth timescales (an effect which may be solely due to the longer duration of the simulations). The migration-only simulations show little variation with migration timescale, and regions in which 10\% of resolved planetesimals collided are found at greater distance from the star, about 4\,au. The migrating giant planets have scattered and shepherded planetesimals into the inner regions, increasing the planetesimal surface density and substantially enhancing eccentricities and, hence, impact velocities. {The migration-induced wall of planetesimals reaches most of the way to the inner edge of our initial planetesimal disk in the migration simulations, thus we expect the collision numbers would continue to rise with decreasing heliocentric distance if we included more of the inner disk.}

The Grand Tack scenario shows the highest collision fraction in the inner disk (Figure \ref{f:FracColl}), exceeding 10\% interior to $\sim$4.5\,au in most simulations (recall that Jupiter began at 3.5\,au in these simulations). The fraction of resolved planetesimals experiencing collisions reaches almost 100\% at the inner edge of the modeled disk, 1\,au, in the Grand Tack simulations. {Again,} we expect the collision numbers would be larger had we included more of the disk interior to 1\,au.

The Grand Tack simulations show little variation in collision fraction with growth and migration timescale, the significant difference is between simulations with the MMSN gas disk compared to the hydrodynamic-derived gas (HG). The HG disk has a higher gas surface density across most of the modeled regions causing greater aerodynamic drag on planetesimals than the MMSN gas disk. The greater gas drag causes stronger, more rapid decay of eccentricities and inclinations which causes fewer planetesimals to be scattered beyond the inner edge of the modeled region. With more planetesimals in the modeled region of the disk and lower inclinations, the collision probability is higher, and the HG disk simulations generally show greater numbers of collisions than their MMSN counterparts.

Figure \ref{f:FracSurvive} shows the fraction of planetesimals that survived to the end of the simulations, remaining bound and within the range of initial disk radii. 
Many of these planetesimals were scattered or shepherded away from their individual starting locations. The majority of planetesimals survived for the duration of the simulations in the growth cases, but more were lost (accreted, disrupted or ejected) in the simulations that included migration (see Table \ref{t:summary}). In growth simulations, approximately 60--80\% of planetesimals that began close to the seeds of the giant planets were ejected or destroyed, planetesimals that began at least 1\,au away from a planet seed almost all survived. In simulations with migration of the giant planets, 50--95\% of planetesimals that began in the regions the giant planets migrated through were lost. Considering the fractions of planetesimals that survived and the fractions that collided together, we see that collisions are not wholly responsible for this planetesimal loss. Ejection of planetesimals via gravitational interactions is also common. 

Figures \ref{f:growthae}, \ref{f:migae}, and \ref{f:GTae} reveal that the planetesimals that started close to, or in the path of the giant planets are transported away from their starting locations. Those planetesimals that began in these regions and survived to the end of the simulation were delivered to other regions of the disk (see Appendix \ref{a:mixing}). Giant planet growth and migration can, thus, aid mixing of chemically distinct regions of the protoplanetary disk \citep[e.g.][]{Walsh11,Raymond17}.

Resolved planetesimal survival rates are lower in the Grand Tack models (Figure \ref{f:FracSurvive}). As the Grand Tack is the most dynamically excited set of scenarios, this is not unexpected. The larger Keplerian orbital velocities and surface densities across the range of disk radii studied in these simulations, together with the asynchronous migration followed by outward migration, are likely responsible for this difference. The higher impact velocities and rates closer to the central star lead to more destructive collisions {than in the migration-only simulations}. {The asynchronous migration of Jupiter and Saturn, followed by outward migration} likely results in greater ejection (scattering) of planetesimals {compared with the migration-only scenarios}. The Grand Tack models all ended with fewer than 50\% of the original planetesimals remaining bound and in the modeled region of the disk.

\subsection{Planetesimal implantation}

\begin{figure}
\includegraphics[width=\columnwidth]{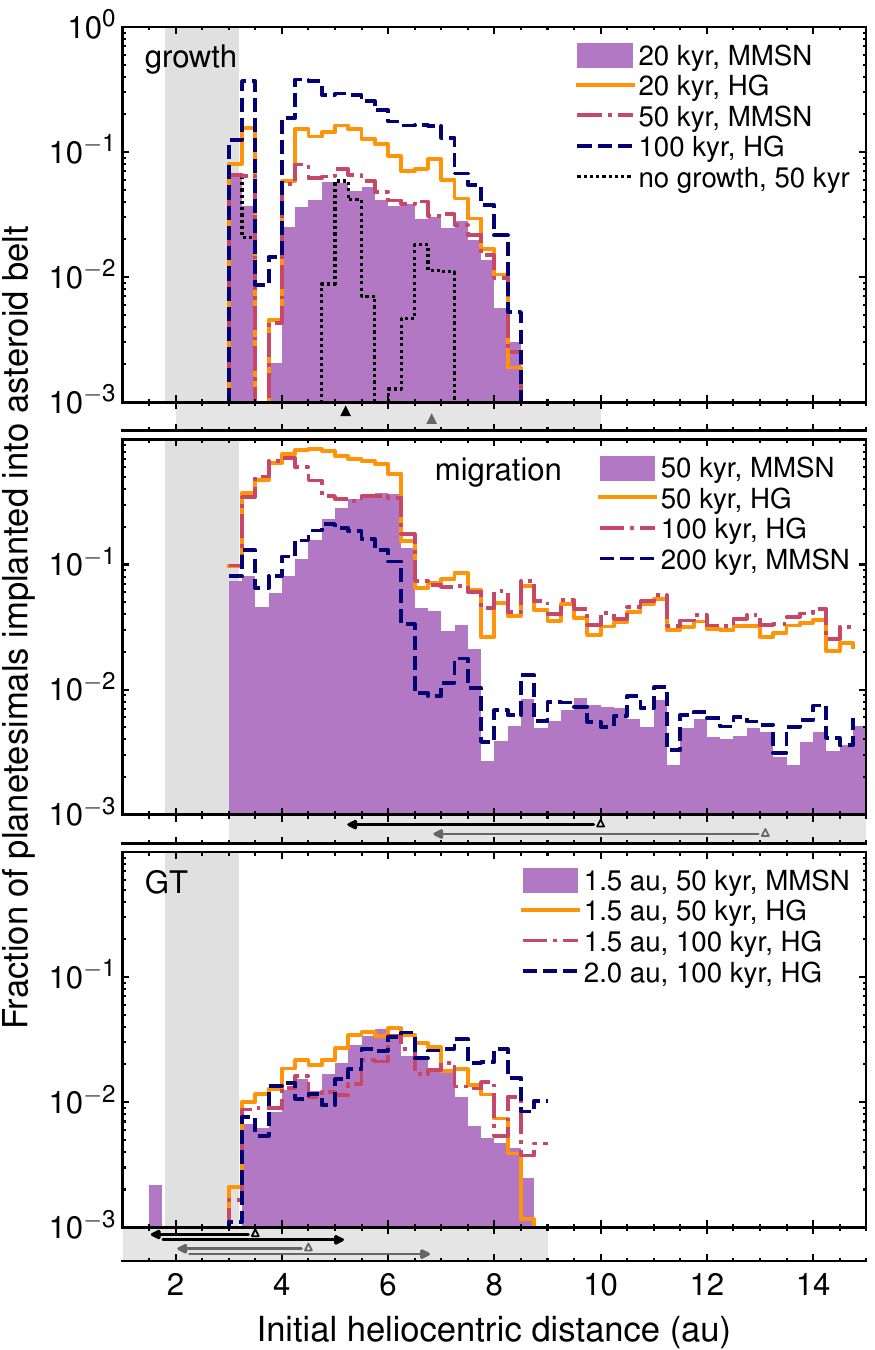}
    \caption{Fraction of resolved planetesimals {from initial locations outside the asteroid belt that were} implanted into the belt (those that have a semi-major axis between 1.8 and 3.2\,au at the end of the simulation). The implantation fractions are averaged across similar simulations. The top panel shows the results from in-situ growth simulations, the middle panel migration-only simulations, and the bottom panel Grand Tack simulations. The area below each panel indicates the initial location and migration of Jupiter (black) and Saturn (grey) as in Figure \ref{f:FracColl}. Note that the solid histogram is used for easier comparison and generally shows the lowest values, it should not be considered the nominal case.}
    \label{f:FracImplant}
\end{figure}

Finally, we examine the effects of giant planet growth and migration on the asteroid belt. The fraction of resolved planetesimals {that began in each region of the disk that were} implanted into the asteroid belt (defined here as semi-major axis between 1.8 and 3.2\,au) is shown in Figure \ref{f:FracImplant}. Note that the implanted planetesimals at the end of these simulations are, by definition, `survivors'. We did not account for planetesimals that would have formed after the start of the simulation, nor did we account for further collisional processing or giant planet evolution after the end of our simulations.

We see broadly similar results to those shown by \citet{Raymond17}, though with generally lower implantation efficiencies {(compare figure 10 in \citeauthor{Raymond17})}. The reduction in implantation is likely due to collisions destroying planetesimals or altering their orbits, and planetesimal-planetesimal interactions causing greater scattering into other regions of the disk, or ejection. Implantation in the migration-only simulations is most efficient for planetesimals with initial heliocentric radii between 4 and 6\,au, as found by \citet{Raymond17}. However, the implantation efficiency is lower and the source regions of implanted planetesimals are different in the Grand Tack simulations. Recall that Jupiter and Saturn begin at 3.5\,au and 4.5\,au in these Grand Tack simulations, causing substantial outward scattering of planetesimals that began between 4 and 5\,au.

In the in-situ growth and migration-only simulations the higher gas density of the hydrodynamic-derived disk (see Figure \ref{f:surfden}) leads to significantly greater implantation into the asteroid belt, due to increased damping and circularization of the orbits instead of scattering into the terrestrial planet region. The Grand Tack simulations, on the other hand, show remarkably consistent, lower implantation fractions from throughout the disk. In all cases, the resolved planetesimal mass implanted into the asteroid belt and the mass of surviving planetesimals from the initial belt region were both much greater than the modern belt mass (see Table \ref{t:belt}).

\begin{table*}
    \centering
    \caption{Summary of asteroid belt region results for all simulation categories. The average fractions of planetesimals that began in the asteroid belt region (1.8--3.2\,au) and underwent at least one collision (collided), and those that began in the belt region and were present in the same region at the end of the simulation (survived) are given. The final two columns give the average mass of planetesimals that survived inside the belt region and the mass implanted into the belt region from outside in multiples of the modern asteroid belt mass, M$_\mathrm{belt} = 5 \times 10^{-4} \,$\mearth \citep{DeMeo13}. The initial disk radii in the growth and migration simulations did not cover the entire asteroid belt region. Note that these values are averages across several randomized initial conditions and simulation resolutions, that have small variations in the initial numbers of planetesimals in the asteroid belt.
    \label{t:belt}}
    \begin{tabular}{lrccrr}
    \hline
                    &          Growth/migration time          & \multicolumn{2}{c}{Fraction of belt planetesimals} & Surviving mass & Implanted mass \\
    Simulation type & scale (kyr)  &  collided     & survived  & (M$_\mathrm{belt}$)  & (M$_\mathrm{belt}$)\\
    \hline
    growth, MMSN        & 20    & 0.042  & 0.97  & 2590$^*$ & 259 \\
    growth, HG          & 20    & 0.044  & 0.97  & 2570$^*$ & 658 \\
    growth, MMSN        & 50    & 0.062  & 0.93  & 2500$^*$ & 368 \\
    growth, MMSN        & 100   & 0.095  & 0.90  & 2470$^*$ & 495 \\
    growth, HG          & 100   & 0.164  & 0.92  & 2490$^*$ & 1470 \\
    \hline
    no growth, MMSN       & --   & 0.071 &  0.93 & 2500$^*$ &  61 \\
    no growth, HG         & --   & 0.070 &  0.93 & 2510$^*$ &  79 \\
    \hline
    migration, MMSN     & 50    & 0.072  & 0.81  & 588$^\dag$ & 2190 \\
    migration, HG       & 50    & 0.097  & 0.82  & 594$^\dag$ & 7260 \\
    migration, MMSN     & 100   & 0.082  & 0.79  & 576$^\dag$ & 1950 \\
    migration, HG       & 100   & 0.098  & 0.81  & 586$^\dag$ & 5400 \\
    migration, MMSN     & 200   & 0.099  & 0.78  & 567$^\dag$ & 1590 \\
    \hline
    GT 1.5\,au, MMSN    & 50    & 0.187  & 0.005  & 28 & 262 \\
    GT 1.5\,au, HG      & 50    & 0.474  & 0.004  & 26 & 334 \\
    GT 2.0\,au, HG      & 50    & 0.376  & 0.004  & 22 & 375 \\
    GT 1.5\,au, MMSN    & 100   & 0.170  & 0.002  & 11 & 228 \\
    GT 1.5\,au, HG      & 100   & 0.434  & 0.002  & 10 & 227 \\
    GT 2.0\,au, HG      & 100   & 0.439  & 0.002  & 14 & 295 \\
    \hline
    \end{tabular}
    \tablecomments{ $^*$\,mass of surviving planetesimals may be underestimated as the inner 0.2\,au of the belt region was not included in the initial disk. Implanted mass may be overestimated as additional collisional losses would likely occur if the interior disk was present.\\
    $^\dag$\,mass of surviving planetesimals is likely a substantial underestimate as the majority of the belt region was not included in the initial disk (migration simulations had no planetesimals interior to 3\,au at the start). Implanted mass may be overestimated as additional collisional losses would likely occur if the interior disk was present.}
\end{table*}
%

%%%%%%%%%%%%%%%%%%% DISCUSSION %%%%%%%%%%%%%%%%%%%%

\section{Discussion}

\subsection{When and where?}

We have seen that growth and migration of giant planets can have a substantial effect on the population of nearby planetesimals. The timings of the growth and migration are, however, uncertain. The simulation types in this work represent different times or possible evolutionary histories for our solar system.

The cores of the giant planets must take some time to grow large enough for rapid gas accretion to begin, but this timescale is uncertain and depends on how and where they accrete \citep[e.g.][]{Lambrechts12}. It is thought that Jupiter (or its core) was able to spatially separate condensed materials in the inner and outer solar system by about 1\,Myr after the start of the solar system (as measured by calcium-aluminum-rich inclusions, CAIs) in order to produce the meteoritic isotope dichotomy \citep[e.g.][]{Scott18}, but other explanations are possible. Disk evaporation may have occurred first in the inner solar system before progressing outward, rather than the simple uniform dissipation used in our models. Jupiter and Saturn must have reached (near) their final masses by the time the nebular gas disk dissipated in their locations. The typical disk lifetime is 3--5\,Myr \citep{Ribas14}. The actual timing of gas dissipation in our solar system is also uncertain, though there is evidence from magnetic fields preserved in meteorites that suggests the gas disk had dissipated by $\sim$4\,Myr after the formation of CAIs \citep{Wang17}.

The timing of planetesimal formation is somewhat better constrained. Some planetesimals formed within the first million years of solar system evolution \citep[e.g.][]{Kleine05,Schersten06}, likely before the giant planets finished growing. The peak of chondrule formation occurs at 2--3\,Myr \citep[e.g.][]{Scott14}, and likely overlaps with the period of giant planet growth and migration. Chondrite parent bodies are expected to have formed contemporaneously with chondrules. Thus, planetesimals would likely still have been forming during the time period we modeled -- a process we did not account for. The unusual {chondrules found in the young, metal-rich CB chondrites} formed at $\sim$4.5\,Myr after CAIs \citep{Krot05} and are believed to be the result of a high-velocity impact. The formation of the CB chondrules may have occurred beyond the end of the time period we modeled, but could coincide with giant planet accretion if it occurred late. 

We therefore expect a population of planetesimals to exist at the start of our simulations, and more to form during, and possibly after, the periods of excitation caused by the growth and migration of the gas giants. The true mass of solid material (planetesimals and dust grains) that existed in the solar system's protoplanetary disk is unknown. The efficiency with which the solid mass in the disk was converted into planetesimals is also not known. We began our simulations with approximately the full MMSN mass of solid material in planetesimals, which we expect to be an overestimate for the time period we consider. Even if the planetesimal formation process was very inefficient, some planetesimals must have formed. Whether the growth of the terrestrial planets was dominated by pebble accretion or planetesimal accretion, planetesimal--planetesimal collisions would have occurred during the time periods we modeled.

We have examined only a small subset of the possible evolutionary histories. Jupiter and Saturn must of course grow at larger heliocentric distances if they migrate inwards to reach their modern locations. Growth further from the Sun would have much less of an effect on the inner disk and the asteroid belt, but requires inward migration which has a larger effect than growth alone. {Figure \ref{f:FracColl} shows that more collisions occur in the inner disk as a result of migration than as a result of growth, even when the growth occurs closer to the star.} The simulations in this work provide illustrative examples of the excitation caused by growth and migration. This work demonstrates that while the effects of giant planet evolution are spatially and temporally restricted, they can be very significant. The times and locations of dynamical excitations and collisions induced by giant planet growth and migration are tied to the time and location of the formation of Jupiter (and Saturn) and the dissipation of the nebular gas disk.

\subsection{The outcomes of high velocity collisions}

We found that giant planet migration induces high velocity collisions between planetesimals \citep[see also][]{Turrini12,Johnson16}. The reader should recall that the collisions found in our simulations occur between planetesimals embedded in the nebular gas disk. It has commonly been assumed that aerodynamic drag from the nebula would result in all planetesimal collisions being low velocity events. Our results indicate that this assumption is invalid. High velocity impacts between planetesimals do occur during the nebular phase of protoplanetary disk evolution. Such collisions are prevalent in the path of migrating giant planets and should not be ignored in planetesimal dynamics or planet accretion models.

The majority of the collisions between planetesimals occurring during giant planet migration {($>$70\%)} are sufficiently high velocity to cause shock-vaporization of water ice, and several percent {(3--18\%)} are fast enough to induce vaporization of forsterite \citep{Davies20}. Vaporization of planetesimal constituents has not generally been considered important during planet formation \citep[see][for a {noteworthy} exception]{Hood09}, but the prevalence of such high velocity impacts makes it a key phenomenon that must be explored in more detail.

Unlike in the case of giant impacts, which are known to cause substantial vaporization \citep[e.g.][]{Lock17,Carter20}, the vapor produced in planetesimal--planetesimal impacts will likely not be gravitationally bound to the remnants of the disrupted planetesimals. Vapor produced by an impact will expand rapidly, and can reach velocities that exceed the escape velocity of planetesimals. Vaporized planetesimal constituents may, thus, more readily escape from the colliding bodies. Separation of impact produced vapor could have major chemical consequences for growing terrestrial planets if it allows some chemical fractionation to occur \citep[see][Davies et al, in prep.]{Davies19,Davies20}.

An impact that is fast enough to cause vaporization (melting) does not generally cause vaporization (melting) of the entire impacting mass. The shock pressures decay rapidly away from the impact site \citep{LeinhardtStewart09,Davison10}. Close to the critical impact velocity required for vaporization (melting) only a small portion of the body will reach pressures sufficient to cause impact-induced vaporization or melting. Thus, the large numbers of high velocity collisions we find do not imply vast masses of melt should be present in the meteorite record.

In the simulations presented here, only a few percent of the planetesimals implanted into the asteroid belt have experienced collisions. However, production of even a small amount of vapor can have a huge effect on collision outcomes \citep{Hood09,Carter19,StewartLPSC19b,StewartLPSC19a}, particularly when these impacts occur during the protoplanetary disk's nebula phase. Shock-produced vapor will expand by many orders of magnitude in volume as it releases to the ambient pressure of the nebula \citep{Davies20}. The rapidly expanding bubble of vaporized planetesimal fragments is embedded within the nebular gas, and will gravitationally collapse once the internal pressure drops below the pressure of the nebula \citep{StewartLPSC19a}, potentially causing the formation of new planetesimals \citep{Carter19}. The consequences of these high velocity collisions require detailed further investigation.

It should also be noted that the collision model used in this work (EDACM) does not account for vaporization during planetesimal collisions. The debris escaping from impacts likely includes a mixture of shock-processed droplets and dust as well as planetesimal fragments. Smaller particles in this ejecta could be rapidly lost from the system due to aerodynamic drag, or experience high velocity collisions with larger planetesimals, and thus would not have reaccreted as efficiently as our simple prescription allows.  Changes to assumed collision outcomes due to vaporization of planetesimal constituents could have some effect on our results, potentially causing the formation of new aggregate planetesimals \citep{Carter19} that could also alter the collision, survival, and implantation statistics. High velocity, vaporizing planetesimal impacts require further study, and the development of new scaling laws to account for these phenomena.

\subsection{Orbit excitation}\label{s:orbexcite}

A noticeable difference between the results of this work and that of \citet{Raymond17} is the higher eccentricities seen for a subset of planetesimals across the disk. Collisions themselves play a role in exciting planetesimal orbits, but it is the gravitational interaction between planetesimals that is largely responsible for these higher eccentricities. Care should be taken when drawing inferences from simulations of planetesimal dynamics and planet formation in which planetesimal-planetesimal interactions are ignored.

\begin{figure*}
%\begin{interactive}{animation}{F10_e_i_grow_mig_mmsn.mp4}
\includegraphics[width=\textwidth]{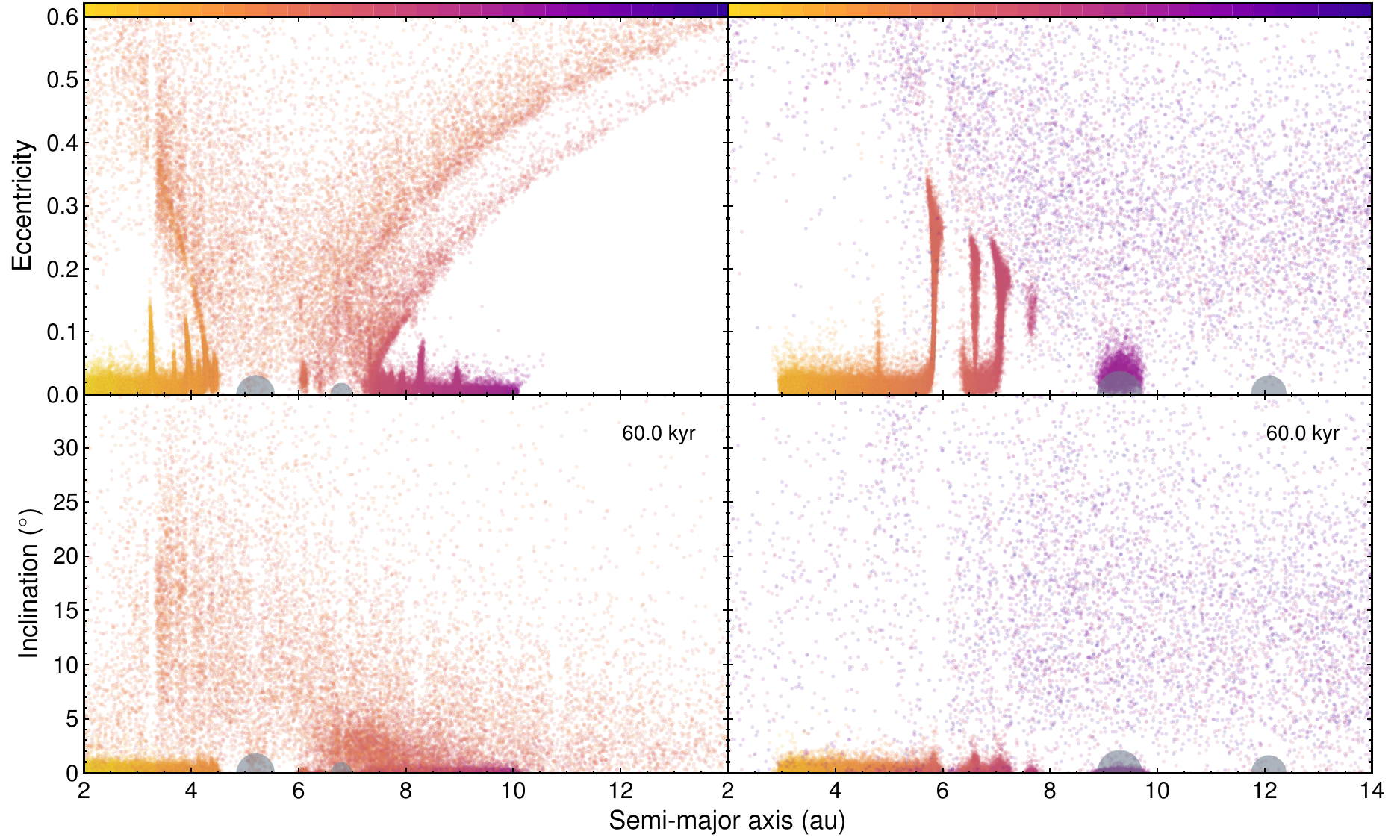}
%\end{interactive}
    \caption{Eccentricity (top) and inclination (bottom) distributions for example growth (left, growthis1e5\_100k\_1) and migration (right, mig10in1e5\_100k\_1) simulations with 100\,kyr growth/migration timescales and MMSN gas disks. Planetesimals are colored according to their average composition as a function of heliocentric distance, Jupiter and Saturn are shown as grey circles. The bodies are shown with partial transparency to highlight densely populated regions. Scattered planetesimals with high eccentricities also have large inclinations. Planetesimals excited by mean motion resonances, on the other hand, retain low inclinations. The high eccentricity ($>\sim$0.1) planetesimals in the growth simulation (mostly orange and pink) also tend to have high inclinations ($>\sim$2$^\circ$), particularly interior to the orbits of the giant planets. In the migration simulations this is also true for the scattered, very high eccentricity planetesimals (mostly purple), but the densely populated resonant walls (orange and pink) retain low inclinations despite their high eccentricities. An animation of this figure is available, it shows eccentricities and inclinations of the planetesimals vs semi-major axis for the first 200\,kyr of evolution. The duration of the video is 25 seconds.}
    \label{f:inclination}
\end{figure*}
We have seen that giant planet migration leads to more collisions than giant planet growth alone. While gas giant growth causes substantial scattering of nearby planetesimals, migration can affect many more bodies, shepherding planetesimals inwards (or outwards) in resonant walls \citep[see also][]{Tanaka99,Fogg05,Mandell07,Carter15}. Planetesimals in these walls encounter the near-circular orbiting planetesimals ahead of them, leading to high impact velocities, especially closer to the star, where the disk surface density is greater.

A somewhat surprising result of our work is that the growth of the gas giants has little effect on impact velocities, with velocity distributions almost identical to those from our control simulations (see Figure \ref{f:velhistcumu}). The planetesimals with high eccentricities induced by giant planet growth also tend to have large inclinations (see Figure \ref{f:inclination}, left panels), and so there are few orbit crossings, and little chance for the excited planetesimals to collide with the planetesimals remaining on circular orbits. The mean motion resonances with the giant planets, on the other hand, produce little inclination excitation (Figure \ref{f:inclination}). The eccentric walls of planetesimals therefore retain low inclinations, and thus have much greater probabilities to interact with planetesimals on near circular orbits, giving rise to high velocity collisions. Giant planet migration thus leads to more collisions, and higher impact velocities, than giant planet growth alone.

\subsection{Implantation of meteorite parent bodies}

We find planetesimal implantation into the asteroid belt is largely consistent with the work of \citet{Raymond17}, although with reduced implantation efficiency due to loss of planetesimals via disruptive collisions. It is reassuring that we find similar results using a different code, with different starting locations for Saturn, and when planetesimal collisions are accounted for. We have used simplified prescriptions for the growth and migration of the gas giant planets; more realistic prescriptions for their evolution may affect our findings, though we expect only small deviations from the general results. Comparison with the control simulations further confirms that growth and/or migration are key in implantation and scattering. 

Of the resolved planetesimals implanted into the asteroid belt, typically 1--5\% experienced collisional processing during the simulations, with the higher percentages occurring for the migration and Grand Tack scenarios. The fraction of planetesimals that underwent collisions was similar for those planetesimals that originated in the asteroid belt and remained in the belt region at the end of the simulations, but reached as high as 10--14\% in some cases. The numbers of collisionally processed objects would likely have been much higher if bodies smaller than our minimum mass limit were considered. The resolved planetesimals that were implanted had also accreted some mass, either from accretive collisions or accretion of unresolved debris. This accreted mass accounts for only $\sim$5\% of the mass in the asteroid belt. The majority of the implanted mass is from bodies that existed at the start of the simulations, though we have likely overestimated the implanted mass (see below).

We found that the inward migration of the giant planets in the Grand Tack causes extensive collisional processing of planetesimals that began in the asteroid belt region (Figure \ref{f:FracColl} and Table \ref{t:belt}). Jupiter's migration through the region of the asteroid belt shepherds the majority of the surviving initial belt into the inner solar system (Figure \ref{f:GTae}). The asteroid belt is, thus, substantially depopulated by the Grand Tack. Less than 1\% of the original asteroid belt planetesimals remain at the end of our simulations (see Table \ref{t:belt}). The outward migration partially replenishes the asteroid belt with a mixture of planetesimals from differing heliocentric radii (Figure \ref{f:FracImplant}; see also \citealt{Walsh11,Walsh12}).

\citet{Raymond17} found implantation was generally much too efficient, emplacing too much mass in the asteroid belt. Collisional losses help to reduce the scale of this problem, but are not sufficient to overcome the over-implantation of mass into the belt in our simulations (see Table \ref{t:summary}). Later collisional processing of the asteroid belt, after the dispersal of the nebular gas, would likely lead to some mass loss. \citet{Raymond17} discussed later dynamical clearing \citep[e.g.][]{Clement18}, and fewer initial planetesimals as possible mechanisms to reduce the implanted mass. They also suggested that lower gas disk surface densities would reduce the efficiency of implantation. We observe significantly less efficient implantation for the MMSN gas disk than for the hydrodynamic-derived gas (HG) disk. The lower gas density, and thus reduced aerodynamic drag, of the MMSN disk reduces the implanted fraction in the growth and migration simulations (Figure \ref{f:FracImplant}), though this reduction in gas surface density does not seem to be sufficient to fully explain the over-implantation. The Grand Tack model leads to substantially lower implantation efficiency than the growth or migration simulations (Figure \ref{f:FracImplant} and Table \ref{t:belt}), as expected \citep{Walsh11}.

Implantation efficiency also depends on planetesimal size \citep{Raymond17}. We did not explore planetesimal size directly in our simulations, instead keeping a fixed disk mass. The factor of five variation in number of particles across our simulations has little effect on planetesimal sizes (350--600\,km). \citet{Raymond17} showed that implantation into the asteroid belt is more efficient for smaller planetesimals. The sizes of the planetesimals modeled in this work (350--600\,km) are close to the larger expected `birth sizes' for planetesimals formed via the streaming instability \citep[e.g.][]{Simon16}. We would expect that real primitive planetesimals may have been somewhat smaller, and thus our implantation efficiency is likely an underestimate. Conversely, we began with approximately the full `minimum mass' of planetesimals expected for the MMSN in the modeled region. Since planetesimals would have been forming throughout, and possibly after, the time periods covered by our simulations the mass of planetesimals available to be implanted during giant planet growth and migration would likely have been lower. Hence, our estimates of planetesimal implantation efficiency may be overestimates.

A low fraction of planetesimals from the outer solar system (beyond $\sim$8\,au) experience collisions in the migration scenarios (Figure \ref{f:FracColl}). However, up to 10\% of the planetesimals that originated between 8 and 14\,au can be implanted into the asteroid belt region due to giant planet migration (Figure \ref{f:FracImplant}). A larger fraction of these resolved outer solar system planetesimals collided in the simulations with giant planet growth, but there was very little implantation of planetesimals that began beyond 8\,au into the asteroid belt in the growth scenarios. The outward migration of Saturn in the Grand Tack shepherds walls of planetesimals outwards and scatters other planetesimals inwards to populate the asteroid belt.

\subsection{The role of impacts in chondrule formation}\label{s:chondruleDiscuss}

Similarly to \citet{Johnson16}, we see impact velocities in excess of 18\,km\,s$^{-1}$ in the asteroid belt region in the Grand Tack scenario (though only a small number of collisions). \citet{Johnson16} used these high impact velocities to tie giant planet migration to the formation of CB chondrites via vaporization of the iron cores of planetesimals. We note that the 18\,km\,s$^{-1}$ critical velocity from \citet{Kraus15} is for shock and release of iron initially at standard temperature and pressure to Earth atmospheric pressure (1\,bar). The critical velocity for shock-induced vaporization of the iron cores of planetesimals would be somewhat lower in the ambient pressure of an evaporating solar nebula \citep{Davies19b,Davies20}. We also see such high impact velocities in the region of the asteroid belt in our in-situ growth and migration-only simulations, and we would expect continued impacts at high velocities in the subsequent Myr after Jupiter's outward migration in the Grand Tack model. We therefore conclude that giant planet growth alone could cause impact-induced vaporization of iron and the formation of the rare CB chondrites, and such a model is not specific to the Grand Tack.

While it is generally accepted that collisions play a critical role in the formation of chondrules in CB chondrites, it is under debate whether collisions contribute to the formation of the silicate chondrules found in most carbonaceous, ordinary, and enstatite chondrites \citep[e.g.][]{ConnollyJones16}. Answering the question of whether collisions form chondrules will require the products of collisions to be examined in great detail.

The treatment of unresolved debris in our simulations is simple. Mass in the unresolved bins remains unresolved unless it is reaccreted by planetesimals passing through the bins. Resolved planetesimals on non-circular, bound orbits accrete unresolved debris according to their cross-sectional area, the density of debris in the annulus the they travel through, and their eccentricity \citep{Leinhardt05}. Planetesimals on circular orbits are assumed not to collide with the debris, and the mass accreted increases as the planetesimal eccentricity increases. This simple reaccretion is likely a substantial overestimate for high eccentricity planetesimals. Resolved planetesimals with high eccentricities would possess large velocities relative to the unresolved debris (which is assumed to be on cold, circular orbits). The collisions between the smaller unresolved bodies and the large resolved planetesimals would likely not be accretive, but the outcome will depend strongly on the size of the debris material. Much smaller debris, such as mm-sized droplets or dust, may act as `buck shot', ablating the surface of the large planetesimals, whereas km-scale planetesimals may cause disruption if they impact the larger bodies at high velocity. More sophisticated treatments of this debris and collisions with these small fragments are needed to advance planet accretion models.

During the nebula phase, the small material (both unprocessed and collisional debris) may be continuously aggregated into new planetesimals, either via the same mechanisms that produced the earlier generations of planetesimals, or via collisionally-produced vapor bubbles (see below; \citealt{Carter19,StewartLPSC19b}). Indeed, chondrite formation mostly occurs between 2 and 3\,Myr after the formation of CAIs \citep{Scott14,Desch18}, which overlaps with the expected period of giant planet growth and migration. Recycling of this collisional debris may play a key role in chondrite formation \citep{Lichtenberg18}. The vapor plumes and debris produced by the highly disruptive, vaporizing collisions we see in our simulations, may act to enhance the generation of new planetesimals \citep{Carter19,StewartLPSC19b}.

Inward migration of the giant planets induces collisions in the inner solar system that have impact velocities high enough to cause vaporization of silicates. During the giant planet growth phase, there is a large fraction of collisions across the protoplanetary disk with impact velocities high enough to induce vaporization of water ice. Due to the orders of magnitude expansion in volume of even a small quantity of vapor \citep{Davies20}, such collisions will produce huge vapor cloud structures \citep{StewartLPSC19a}. Since these collisions are necessarily occurring within the protoplanetary nebula, the presence of the nebular gas cannot be ignored. Vapor plumes produced by impacts between planetesimals drive a shock in the nebular gas capable of melting solid particles in the nebula \citep{Hood09,StewartLPSC19a}. The gravitationally collapsing vapor plume would size-sort and collect dust and melt droplets, producing a warm, dense cloud with a mixture of solids similar to chondrites \citep{LockLPSC19,Carter19}. Vaporizing impacts between planetesimals embedded in the nebula may thus be responsible for the formation of chondrules and chondrites \citep{StewartLPSC19b,Carter19}. The products of these collisions can then be scattered into the region of the asteroid belt by the migration of Jupiter and Saturn, or formed in the belt region as a result of collisions caused by giant planet excitation. The growth and migration of Jupiter and Saturn may thus be crucial in the formation (as well as the implantation into the asteroid belt, \citealt{Raymond17}) of the parent bodies of many of the known meteorites.

%%%%%%%%%%%%%%%%%%%% SUMMARY %%%%%%%%%%%%%%%%%%%%%%

\section{Summary}

We have shown that giant planet migration induces a large number of high velocity {(potentially disruptive)} collisions between planetesimals. We have identified several important aspects of planetesimal collisions that are not currently included in planet accretion models{: disruptive collisions, vaporization of planetesimal constituents, the effects of planetesimal debris, the possible production of chondrules, and the potential formation of second-generation planetesimals}. {Extremely energetic collisions make up a small but important fraction of the impact velocity distribution in all the cases we examined.} The classical view, that planetesimal collisions are slow, inconsequential events, is not true during giant planet growth and migration {(nor even in the presence of non-evolving planets)}.

The numbers, impact velocities, and timing of planetesimal-planetesimal collisions depend on the specific growth and migration path of the giant planets. We find that giant planet growth has very little effect on impact velocities compared to a control case {(with embedded planet cores that do not increase in mass)} over the short period of giant planet growth as most of the planetesimals scattered by growing giants do not undergo collisions. Migration, however, greatly increases the number and velocities of collisions. Collisional evolution reduces the efficiency of planetesimal implantation into the asteroid belt compared to that found in other studies, but does not prevent implantation. A small fraction of the largest planetesimals that remain or are implanted into the asteroid belt are processed via collisions.

High velocity planetesimal-planetesimal impacts can induce partial shock-vaporization of both water ice and silicate components of planetesimals. Vapor and debris produced by disruptive, vaporizing collisions may enhance the generation of new planetesimals and may contribute to chondrule formation. The prevalence of high velocity collisions during Jupiter's migration makes impact vaporization in planetesimals an important phenomenon that may have major consequences for meteorite parent bodies. 

Giant planet growth and migration play a major role in sculpting planetary systems. The consequences of vaporizing planetesimal collisions require detailed further study, and development of collision outcome models for impacts within the nebula, and models for the formation of new planetesimals in the aftermath of collisions.

\acknowledgments

{The authors thank the two anonymous reviewers for helpful comments and suggestions that increased the quality and clarity of this manuscript.}
The authors also thank Simon Lock and Bethany Chidester for discussions and comments that substantially improved this manuscript. 
PJC acknowledges financial support from UCOP LFRP and NASA 80NSSC18K0828 (NEXSS).
This research has made use of NASA's Astrophysics Data System.\bigskip\\ 
{Data from the simulations presented in this work are available from \citet{Carter20CollisionData}.}

%PJC would like to thank Sir F Drake for providing emotional support and a limitless supply of fur.\\


\software{PKDGRAV \citep{Richardson00, Stadel01},  
          NumPy \citep{numpy}, 
          matplotlib \citep{matplotlib}.
          }

%\bibliography{references}{}

\bibliographystyle{aasjournal}

\appendix

\setcounter{table}{0}
\renewcommand{\thetable}{A\arabic{table}}

\setcounter{figure}{0}
\renewcommand{\thefigure}{A\arabic{figure}}

\section{Hydrodynamic simulation derived nebular gas surface density profile}\label{a:gas}

In some simulations we adopt an alternative nebular gas profile (labelled `HG') instead of the standard MMSN gas surface density. This alternative gas prescription is based on hydrodynamic simulations from \citet{Morbi07}, and was used in the original Grand Tack models \citep{Walsh11}. This gas profile accounts for the `gap' opened in the disk by the fully-grown Jupiter.

\citet{Raymond17} used a version of this same gas density profile, interpolating between it and a simple smooth power law surface density to introduce the gap as Jupiter and Saturn grew. We adopt a simpler approach and neglect the gradual development of the gap. We expect this to have little effect on our results as the region of the gap is continuously excited by the giant planet cores, and rapidly depleted of planetesimals via scattering.

Our implementation of this hydrodynamic-derived surface density follows the approach from \citet{Carter15}, and uses a simple polynomial fit to the density profile provided by \citet{Walsh11}. We have used a new fit for this work to better account for the region beyond the orbit of Jupiter which was ignored by \citet{Carter15}.

The nebular gas surface density, $\Sigma_\mathrm{g}$, in the HG cases was defined according to,
\begin{equation}\label{e:gas}
    \Sigma_\mathrm{g} = 
    \begin{cases}
    \left(\frac{5.2\,\mathrm{au}}{s_\mathrm{J}}\right) \left( 20930  \left(\frac{s}{s_\mathrm{J}}\right)^3 - 46780  \left({\frac{s}{s_\mathrm{J}}}\right)^2 + 24840 \frac{s}{s_\mathrm{J} } + 687.1 \right) \mathrm{\;g\,cm^{-2}} & s \leq 0.94\, s_\mathrm{J} \\
    
    \left(\frac{5.2\,\mathrm{au}}{s_\mathrm{J}}\right) \left( 193.8  \left(\frac{s}{s_\mathrm{J}}\right)^3 - 2637  \left(\frac{s}{s_\mathrm{J}}\right)^2 + 10950 \frac{s}{s_\mathrm{J}} -10100 \right) \mathrm{\;g\,cm^{-2}} & 1.3\, s_\mathrm{J} \leq s < 5.75\, s_\mathrm{J} \\
    
    \left(\frac{5.2\,\mathrm{au}}{s_\mathrm{J}}\right)\, 90 \mathrm{\;g\,cm^{-2}} & \mathrm{otherwise}
    \end{cases},
\end{equation}
where $s$ is the midplane distance from the star, and $s_\mathrm{J}$ is the $s$ of Jupiter in au. Using this expression the gas (surface) density can be calculated straightforwardly for any particle without the use of a table. {The definition based on $s_\mathrm{J}$ ensures that the gap remains with Jupiter in simulations involving gas giant migration.} The exponential decrease in the surface density with time is applied by multiplying equation \ref{e:gas} by the decay function. The radial profile is assumed not to change as the disk dissipates. The initial surface density profile obtained from this model, with Jupiter at 5.2\,au, is shown with a grey dashed line in Figure \ref{f:surfden}.

\section{Evolution in `no growth' simulation}\label{a:control}

\begin{figure}
\centering
%\begin{interactive}{animation}{FA1_nogrow5e450k_hg1_aeiv_mov.mp4}
\includegraphics[width=0.48\columnwidth]{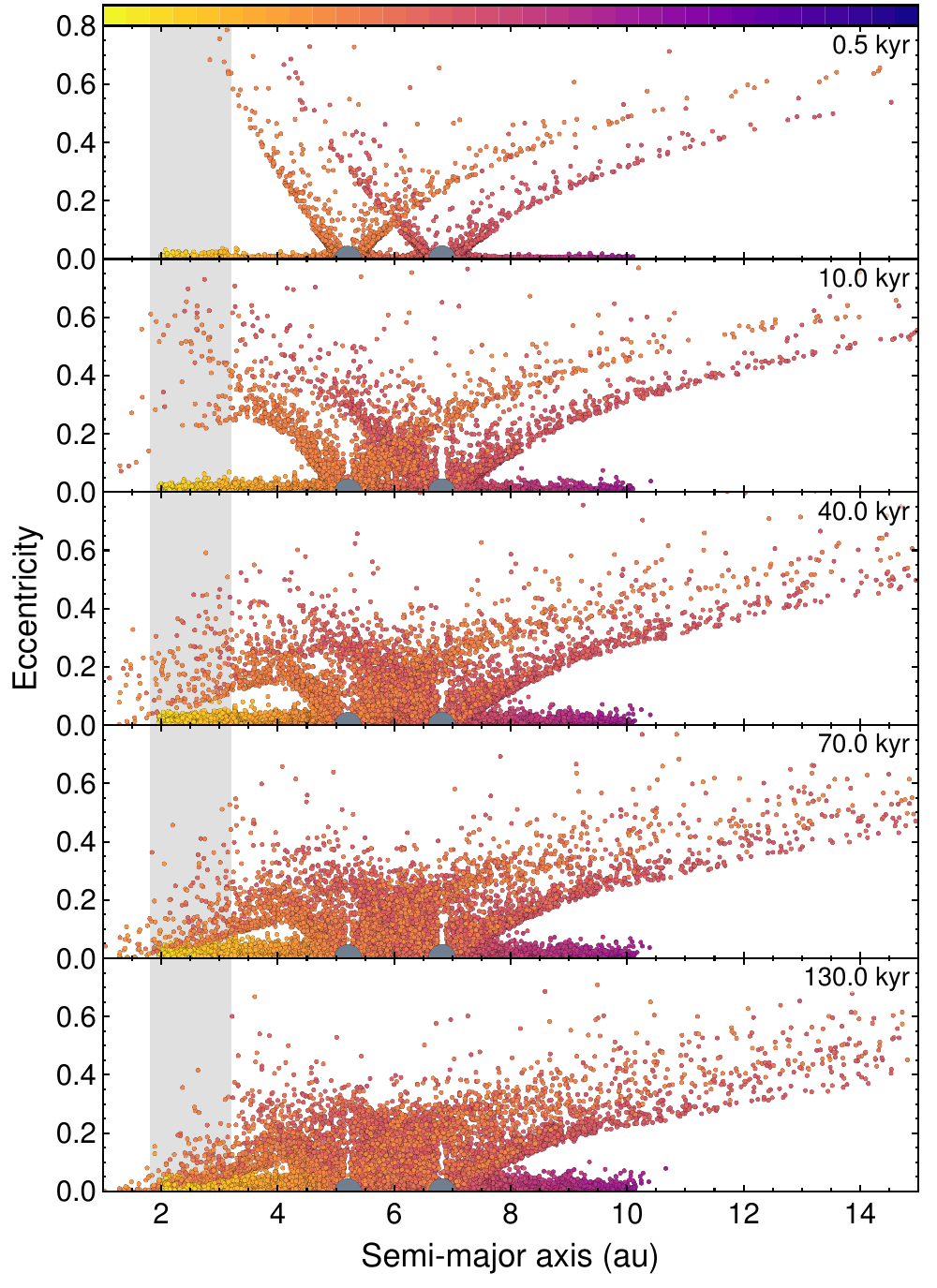}
%\end{interactive}
    \caption{Planetesimal disk eccentricity evolution through five time snapshots in an example control (no growth) simulation (nogrow5e450k\_hg1). Planetesimals are colored according to their average composition as a function of heliocentric distance, proto-Jupiter and proto-Saturn are shown as grey circles. Body sizes are proportional to their mass. 
    An animation of this figure is available, it shows eccentricities and inclinations of the bodies and impact velocities vs semi-major axis for the 130\,kyr duration of the simulation. The duration of the video is 33 seconds.}
    \label{f:nogrowae}
\end{figure}

Figure \ref{f:nogrowae} shows an example of the planetesimal disk eccentricity evolution in a control simulation where the giant planet cores do not grow. At early times, the evolution is very similar to that shown in Figure \ref{f:growthae}. Without growth of the 3\,\mearth cores, however, the eccentricities of planetesimals remain lower than in the growth case, and fewer planetesimals are scattered away from their initial locations.

\section{Further evolution examples}\label{a:examples}

\begin{figure}
\centering
%\begin{interactive}{animation}{FA2_growthis5e4_100k_1_aeiv_mov.mp4}
\includegraphics[width=0.48\columnwidth]{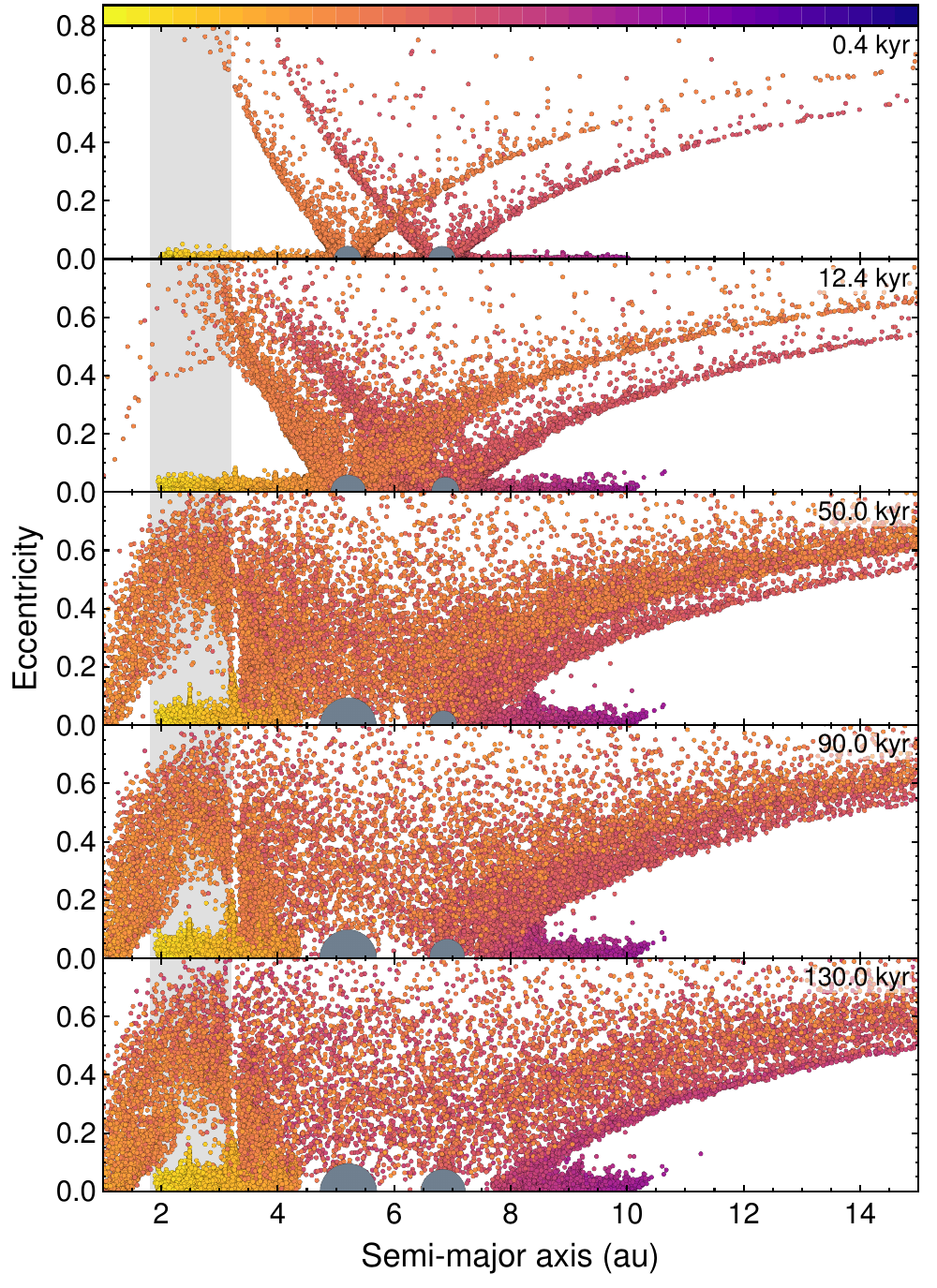}
%\end{interactive}
    \caption{Planetesimal disk eccentricity evolution through five time snapshots in an example in-situ growth simulation with a 50\,kyr growth timescale and MMSN gas disk (growthis5e4\_100k\_1). Planetesimals are colored according to their average composition as a function of heliocentric distance, (proto-)Jupiter and (proto-)Saturn are shown as grey circles. Body sizes are proportional to their mass. The light grey band indicates the asteroid belt region. 
    An animation of this figure is available, it shows eccentricities and inclinations of the bodies and impact velocities vs semi-major axis for the 130\,kyr duration of the simulation. The duration of the video is 33 seconds.}
    \label{f:growthaeA}
\end{figure}
\begin{figure}
\centering
%\begin{interactive}{animation}{FA3_mig10in2e5_100k_1_aeiv_mov.mp4}
\includegraphics[width=0.48\columnwidth]{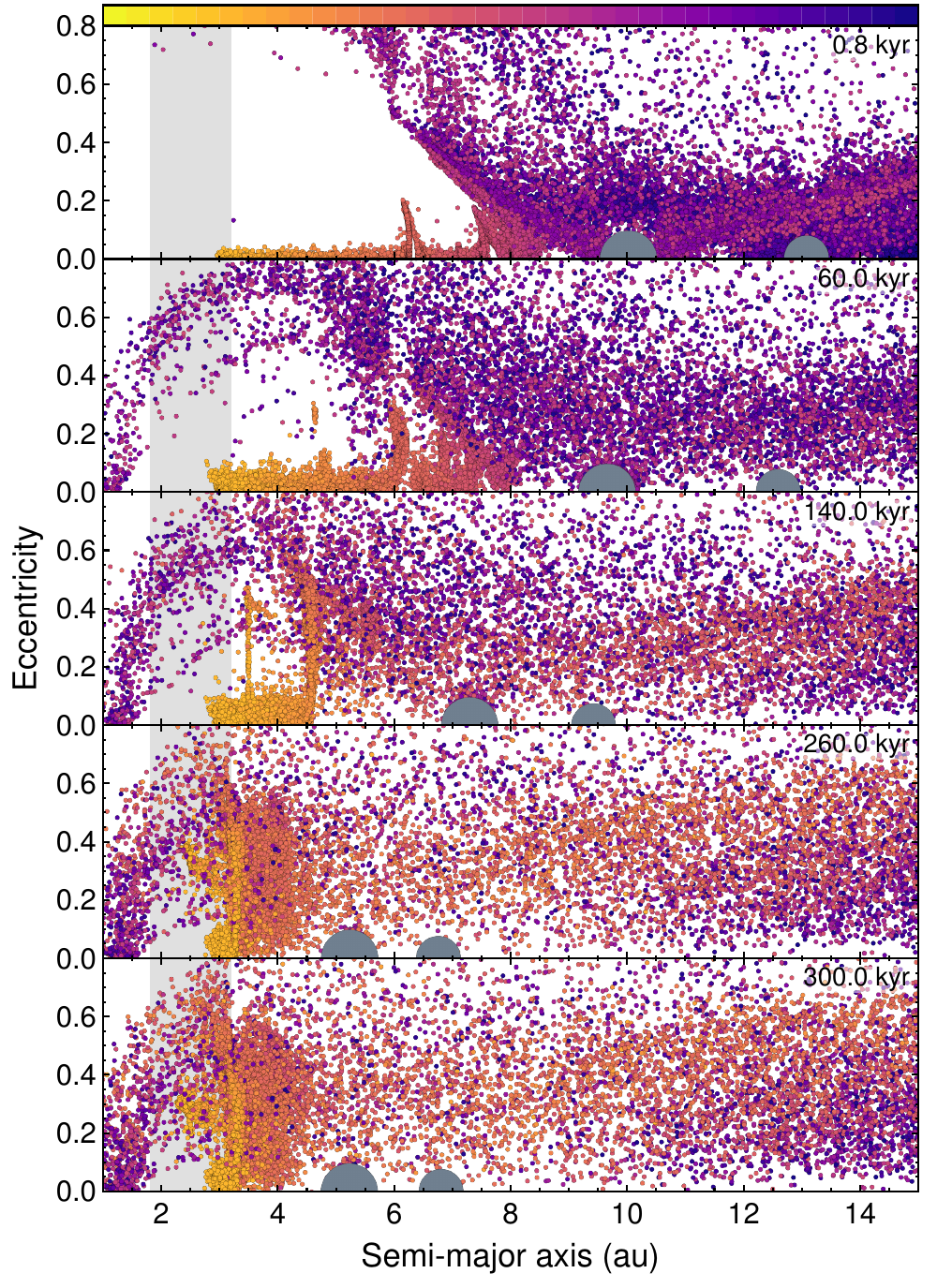}
%\end{interactive}
    \caption{Planetesimal disk eccentricity evolution through five time snapshots in an example migration simulation with a 200\,kyr migration timescale and MMSN gas disk (mig10in2e5\_100k\_1). Planetesimals are colored according to their average composition as a function of heliocentric distance, Jupiter and Saturn are shown as grey circles.  The light grey band indicates the asteroid belt region. 
    An animation of this figure is available, it shows eccentricities and inclinations of the bodies and impact velocities vs semi-major axis for the 300\,kyr duration of the simulation. The duration of the video is 38 seconds.}
    \label{f:migaeA}
\end{figure}
\begin{figure}
\centering
%\begin{interactive}{animation}{FA4_GT15gm5e4_100k_1_aeiv_mov.mp4}
\includegraphics[width=0.48\columnwidth]{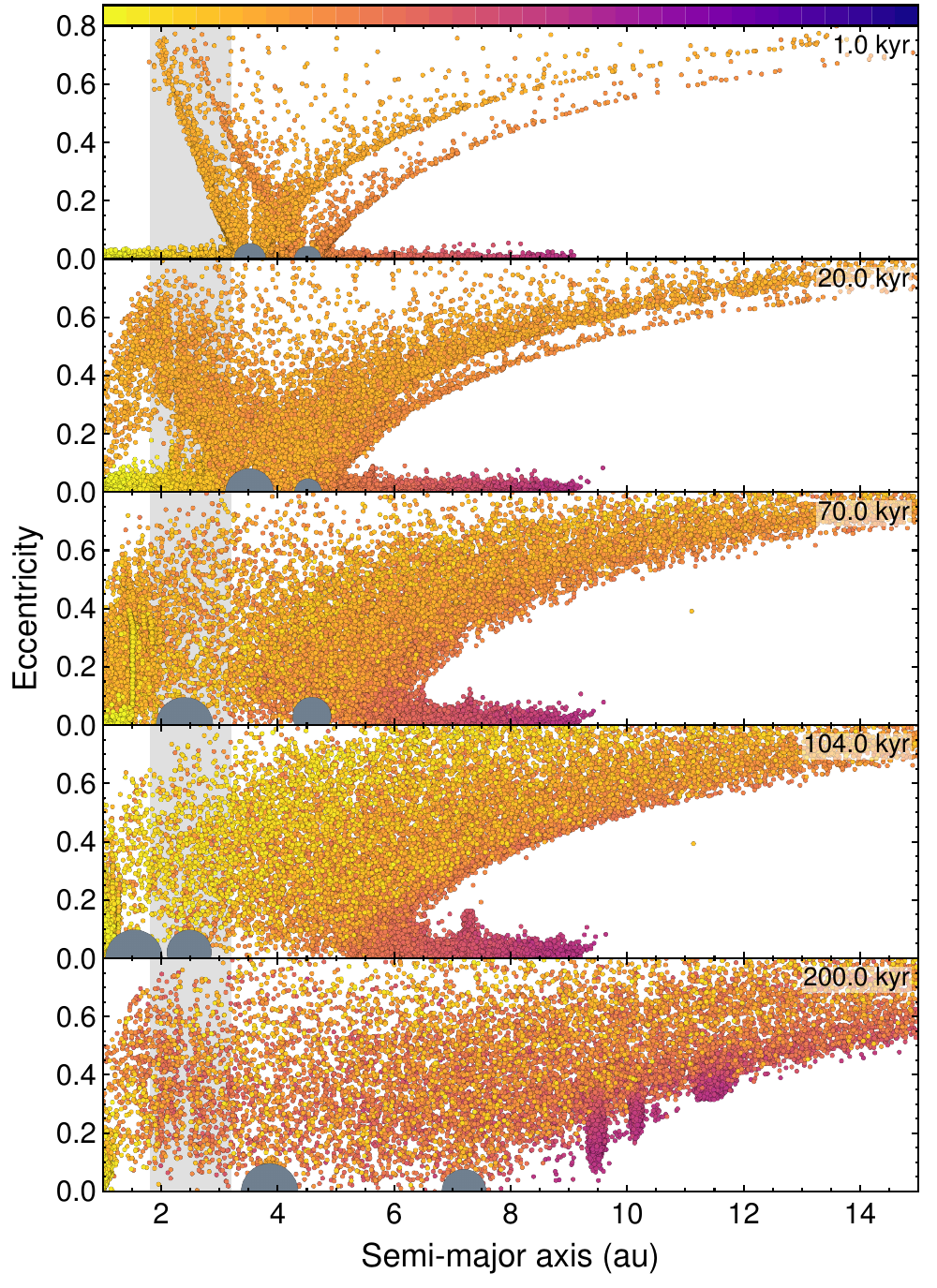}
%\end{interactive}
    \caption{Planetesimal disk eccentricity evolution through five time snapshots in an example Grand Tack simulation with a 50\,kyr evolution timescale and MMSN gas disk (GT15gm5e4\_100k\_1). Planetesimals are colored according to their average composition as a function of heliocentric distance, (proto-)Jupiter and (proto-)Saturn are shown as grey circles. Body sizes are proportional to their mass.  The light grey band indicates the asteroid belt region. 
    An animation of this figure is available, it shows eccentricities and inclinations of the bodies and impact velocities vs semi-major axis for the 200\,kyr duration of the simulation. The duration of the video is 40 seconds.}
    \label{f:GTaeA15}
\end{figure}
\begin{figure}
\centering
%\begin{interactive}{animation}{FA5_GT20gm1e5_50k_hg1_aeiv_mov.mp4}
\includegraphics[width=0.48\columnwidth]{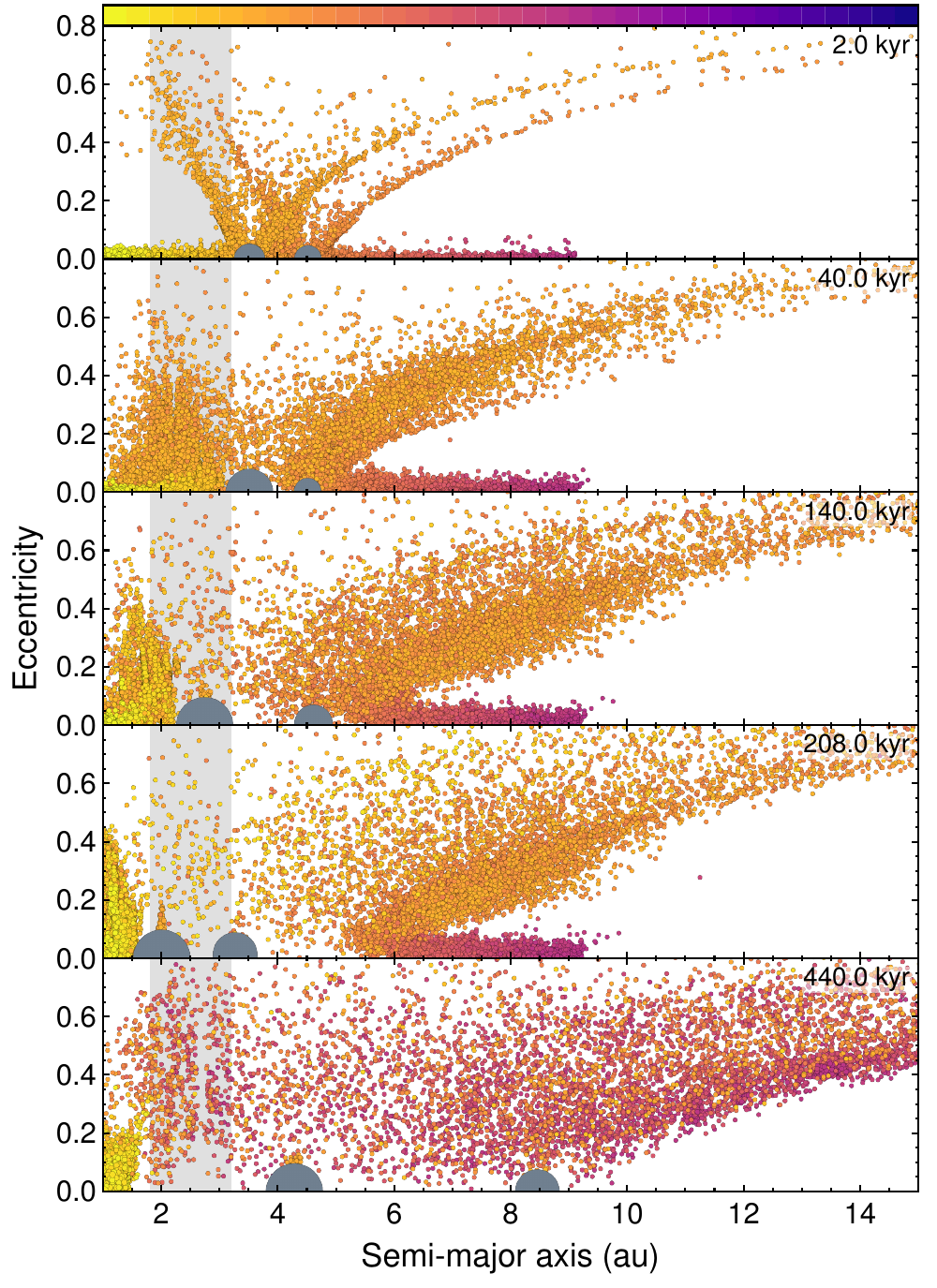}
%\end{interactive}
    \caption{Planetesimal disk eccentricity evolution through five time snapshots in an example Grand Tack simulation with the tack at 2\,au, a 100\,kyr evolution timescale and HG gas disk (GT20gm1e5\_50k\_hg1). Planetesimals are colored according to their average composition as a function of heliocentric distance, (proto-)Jupiter and (proto-)Saturn are shown as grey circles. Body sizes are proportional to their mass.  The light grey band indicates the asteroid belt region. 
    An animation of this figure is available, it shows eccentricities and inclinations of the bodies and impact velocities vs semi-major axis for the 500\,kyr duration of the simulation. The duration of the video is 50 seconds.}
    \label{f:GTaeA20}
\end{figure}

Figures \ref{f:growthaeA}--\ref{f:GTaeA20} show more examples of evolution in the growth, migration and Grand Tack evolution scenarios with different simulation parameters compared to the examples shown in Figures \ref{f:growthae}--\ref{f:GTae}.

{\section{Planetesimal disk mixing}\label{a:mixing}}

\begin{figure}
\centering
\includegraphics[width=0.48\columnwidth]{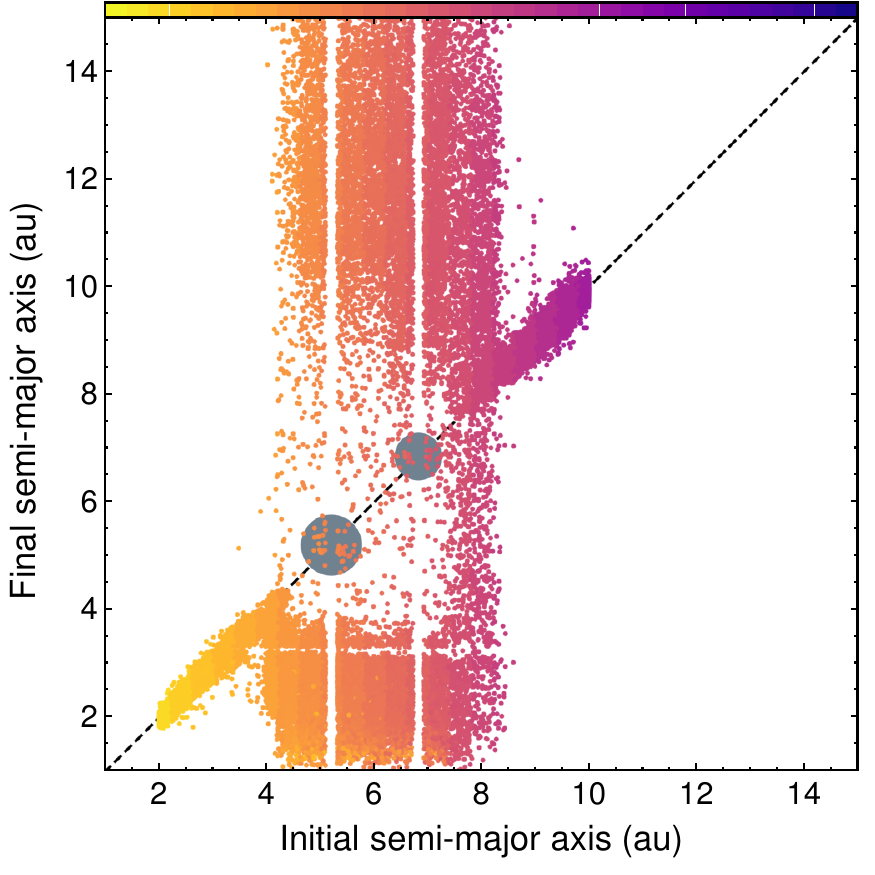}
    \caption{{Final vs initial semi-major axes for bodies in an example growth simulation (growthis1e5\_100k\_hg1; Figure \ref{f:growthae}). Planetesimals are colored according to their average composition as a function of heliocentric distance, proto-Jupiter and proto-Saturn are shown as grey circles. Body sizes are proportional to their mass.}}
    \label{f:growaa}
\end{figure}
\begin{figure}
\centering
\includegraphics[width=0.48\columnwidth]{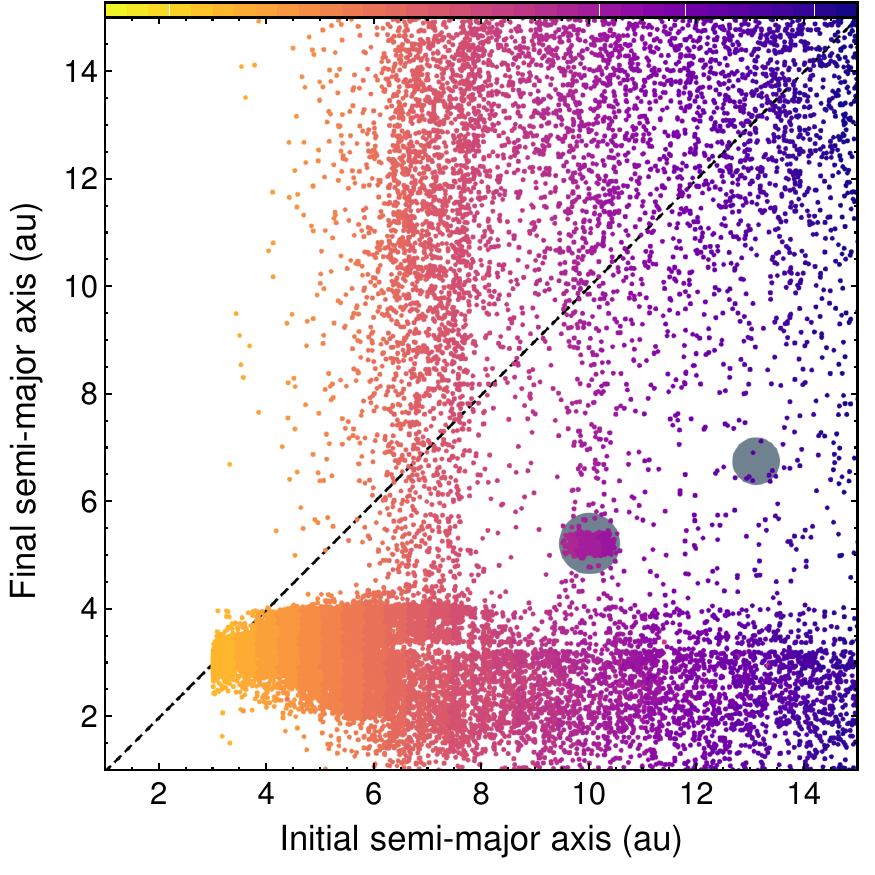}
    \caption{{Final vs initial semi-major axes for bodies in an example migration simulation (mig10in1e5\_100k\_hg1; Figure \ref{f:migae}). Planetesimals are colored according to their average composition as a function of heliocentric distance, proto-Jupiter and proto-Saturn are shown as grey circles. Body sizes are proportional to their mass.}}
    \label{f:migaa}
\end{figure}
\begin{figure}
\centering
\includegraphics[width=0.48\columnwidth]{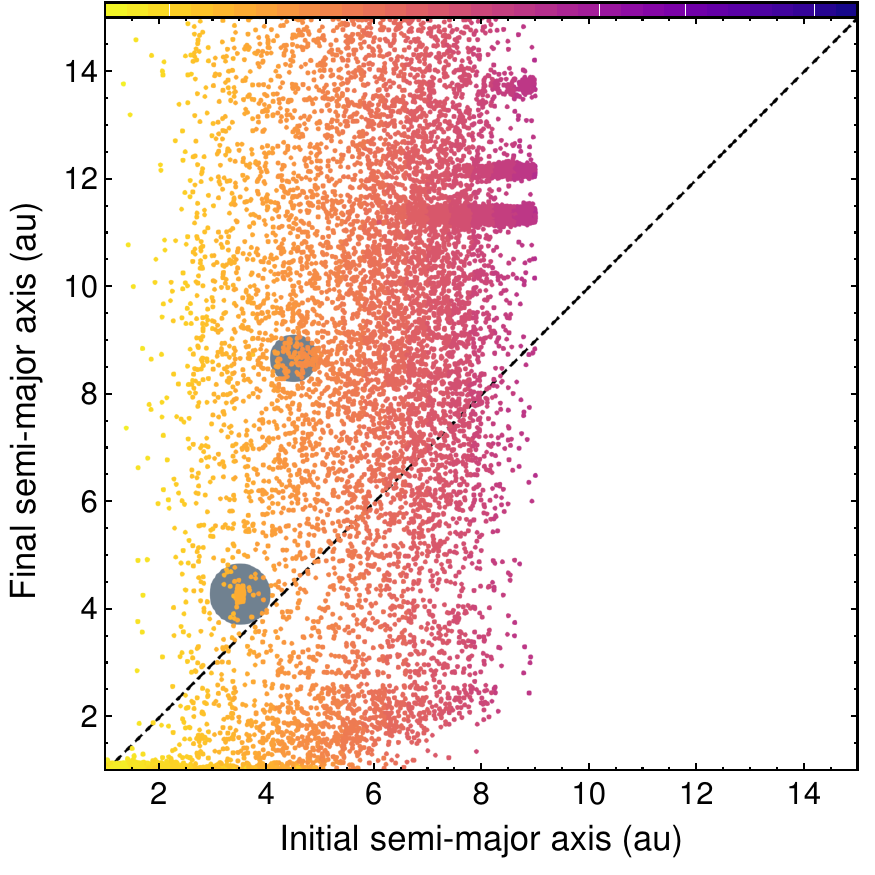}
    \caption{{Final vs initial semi-major axes for bodies in an example Grand Tack simulation (GT15gm5e4\_50k\_hg1; Figure \ref{f:GTae}). Planetesimals are colored according to their average composition as a function of heliocentric distance, proto-Jupiter and proto-Saturn are shown as grey circles. Body sizes are proportional to their mass.}}
    \label{f:GTaa}
\end{figure}
\begin{figure}
\centering
\includegraphics[width=0.48\columnwidth]{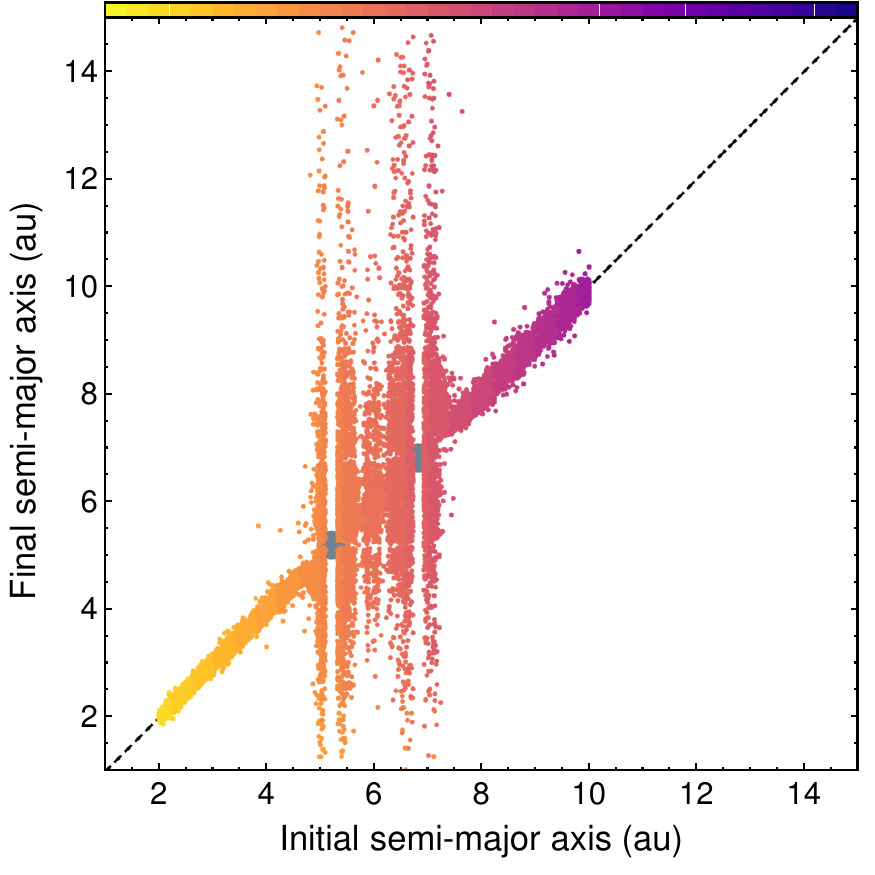}
    \caption{{Final vs initial semi-major axes for bodies in an example control simulation (nogrow5e450k\_hg1; Figure \ref{f:nogrowae}). Planetesimals are colored according to their average composition as a function of heliocentric distance, proto-Jupiter and proto-Saturn are shown as grey circles. Body sizes are proportional to their mass.}}
    \label{f:nogrowaa}
\end{figure}

{Figures \ref{f:growaa}--\ref{f:nogrowaa} show examples of the planetesimals final vs initial semi-major axes. An undisturbed disk would lie on the diagonal black line. The growth simulation (Figure \ref{f:growaa}) shows the planetesimal disk is substantially disturbed and cleared in the vicinity of the growing giant planets, but is almost unaffected further away. The simulations with giant planet migration (Figures \ref{f:migaa} and \ref{f:GTaa}), on the other hand, show that almost the entire simulated disk is disturbed by the migrating giant planets. Much of the interior disk has been shepherded to smaller semi-major axes, and mixed with pink and blue material that began at greater distances from the star.}

\section{Simulation parameter tables}\label{a:tables}

Tables \ref{t:sims}--\ref{t:sims3} give additional parameters for the full set of simulations.

\begin{table*}
\centering
\caption{Summary of additional growth and control simulation parameters. $N$ is the number of planetesimals in the simulation, $\tau_\mathrm{grow}$ is the growth time, $\tau_\mathrm{mig}$ is the migration timescale, $t_\mathrm{start,J/S}$ is the time at which Jupiter/Saturn begin growing, $\Delta t$ is the simulation timestep, $\tau_\mathrm{gas}$ is the gas disk dissipation timescale, $R_\mathrm{init}$ is the starting planetesimal radius, and $M_\mathrm{min}$ is the mass resolution limit.
\label{t:sims}}
\begin{tabular}{lrrrrrrrrrrr}
  \hline
  Name &	$N$ &	$\tau_\mathrm{grow}$	& $\tau_\mathrm{mig}$ 	& $t_\mathrm{start,J}$	& $t_\mathrm{start,S}$	& Disk	& $\Delta t$	& Duration	& $\tau_\mathrm{gas}$ & $R_\mathrm{init}$ & $M_\mathrm{min}$ \\
   & &	(kyr)	& (kyr)		& (kyr)	& (kyr)	&	& (yr)	& (kyr)	& (kyr) & (km) & (\mearth)\\
  \hline
  growthis2e4\_1	    &   20000 &	20	& -	&  5	    & 35	& MMSN	& 0.025	& 60	    & 50 &	599 & 3$\times$10$^{-4}$ \\
  growthis2e4\_2	    &   20000 &	20	& -	&  5     & 35	& MMSN	& 0.025	& 60	    & 50 &	599 & 3$\times$10$^{-4}$ \\
  growthis2e4\_4	    &   20000 &	20	& -	&  5	    & 35	& MMSN	& 0.025	& 60	    & 50 &	599 & 3$\times$10$^{-4}$ \\
  growthis2e4\_5	    &   20000 &	20	& -	&  5	    & 35	& MMSN	& 0.025	& 60	    & 50 &	599 & 3$\times$10$^{-4}$ \\
  growthis2e4\_6	    &   20000 &	20	& -	&  5	    & 35	& MMSN	& 0.025	& 60    & 50 &	599 & 3$\times$10$^{-4}$ \\
  growthis2e4\_7	    &   20000 &	20	& -	&  5	    & 35	& MMSN	& 0.025	& 60	    & 50 &	599 & 3$\times$10$^{-4}$ \\
  growthis2e4\_8	    &   20000 &	20	& -	&  5	    & 35	& MMSN	& 0.025	& 60	    & 50 &	599 & 3$\times$10$^{-4}$ \\
  growthis2e4\_9	    &   20000 &	20	& -	&  5	    & 35	& MMSN	& 0.025	& 60	    & 50 &	599 & 3$\times$10$^{-4}$ \\
  growthis2e4\_10       &   20000 &	20	& -	&  5	    & 35	& MMSN	& 0.025	& 60	    & 50 &	599 & 3$\times$10$^{-4}$ \\
  growthis2e4\_100k\_1	&   100000 & 20	& -	& 5	& 35	& MMSN	& 0.05	& 60	& 50	& 350 & 6$\times$10$^{-5}$ \\
  \hline
  growthis2e4\_hg1      &   20000 &	20	& -	&  5	    & 35	& HG	& 0.025	& 60	    & 50 &	599 & 3$\times$10$^{-4}$ \\
  growthis2e4\_hg2      &	20000 &	20	& -	&  5	    & 35	& HG	& 0.025	& 60	    & 50 &	599 & 3$\times$10$^{-4}$ \\
  growthis2e4\_hg3      &	20000 &	20	& -	&  5	    & 35	& HG	& 0.025	& 60	    & 50 &	599 & 3$\times$10$^{-4}$ \\
  growthis2e4\_hg4      &	20000 &	20	& -	&  5	    & 35	& HG	& 0.025	& 60	    & 50 &	599 & 3$\times$10$^{-4}$ \\
  \hline
  growthis5e4\_1	    &   50000 & 50 &  -	& 10    & 80    & MMSN	& 0.05 &	130 & 100 &	442 & 3$\times$10$^{-4}$ \\
  growthis5e4\_2	    &   50000 & 50	& -	& 10	& 80	& MMSN	& 0.05	& 130 & 100 & 442 & 3$\times$10$^{-4}$ \\
  growthis5e4\_3	    &   50000 & 50	& -	& 10	& 80	& MMSN	& 0.05	& 130 & 100 & 442 & 3$\times$10$^{-4}$ \\
  growthis5e4\_4	    &   50000 & 50	& -	& 10	& 80	& MMSN	& 0.05	& 130 & 100 & 442 & 3$\times$10$^{-4}$ \\
  growthis5e4\_5	    &   50000 & 50	& -	& 10	& 80	& MMSN	& 0.05	& 130 & 100 & 442 & 3$\times$10$^{-4}$ \\
  growthis5e4\_6	    &   50000 & 50	& -	& 10	& 80	& MMSN	& 0.05	& 130 & 100 & 442 & 3$\times$10$^{-4}$ \\
  growthis5e4\_7	&       50000 & 50	& -	& 10	& 80	& MMSN	& 0.05	& 130 & 100 & 442 & 3$\times$10$^{-4}$ \\
  growthis5e4\_8	&       50000 & 50	& -	& 10	& 80	& MMSN	& 0.05	& 130 & 100 & 442 & 3$\times$10$^{-4}$ \\
  growthis5e4\_9	&   50000 & 50 & - & 10	& 80	& MMSN	&   0.05	& 130 & 100 & 442 & 3$\times$10$^{-4}$ \\
  growthis5e4\_10	&   50000 & 50 & - & 10	& 80	& MMSN	&   0.05	& 130 & 100 & 442 & 3$\times$10$^{-4}$ \\
  growthis5e4\_11	&   50000 & 50 & - & 10	& 80	& MMSN	&   0.05	& 130 & 100 & 442 & 3$\times$10$^{-4}$ \\
  growthis5e4\_12	&   50000 & 50 & - & 10	& 80	& MMSN	&   0.05	& 130 & 100 & 442 & 3$\times$10$^{-4}$ \\
  growthis5e4\_t2\_1	&   50000 & 50 & - & 10	& 80	& MMSN	& 0.025	& 130 & 100 & 442 & 3$\times$10$^{-4}$ \\
  growthis5e4\_t2\_2	&   50000 & 50 & - & 10	& 80	& MMSN	& 0.025	& 130 & 100 & 442 & 3$\times$10$^{-4}$ \\
  growthis5e4\_100k\_1	&   100000 & 50	& -	& 10	& 80	& MMSN	& 0.05	& 130	& 100	& 350 & 6$\times$10$^{-5}$ \\
  \hline
  growthis1e5\_1	    &   20000 &	100	& -	&  20	& 200	& MMSN	& 0.025	& 300	& 200 &	599 & 3$\times$10$^{-4}$ \\
  growthis1e5\_2	    &   20000 &	100	& -	&  20	& 200	& MMSN	& 0.025	& 300	& 200 &	599 & 3$\times$10$^{-4}$ \\
  growthis1e5\_3	    &   20000 &	100	& -	&  20	& 200	& MMSN	& 0.025	& 300	& 200 &	599 & 3$\times$10$^{-4}$ \\
  growthis1e5\_4	    &   20000 &	100	& -	&  20	& 200	& MMSN	& 0.025	& 300	& 200 &	599 & 3$\times$10$^{-4}$ \\
  growthis1e5\_100k\_1	&   100000 & 100 & - & 20	& 200	& MMSN	& 0.05	& 300	& 200	& 350 & 3$\times$10$^{-4}$ \\
  growthis1e5\_100k\_2	&   100000 & 100 & - & 20	& 200	& MMSN	& 0.05	& 300	& 200	& 350 & 3$\times$10$^{-4}$ \\
  \hline
  growthis1e5\_100k\_hg1 &  100000 & 100 & - & 20	& 200	& HG	& 0.05	& 300	& 200	& 350 & 6$\times$10$^{-5}$ \\
  growthis1e5\_100k\_hg2 &  100000 & 100 & - & 20	& 200	& HG	& 0.05	& 300	& 200	& 350 & 6$\times$10$^{-5}$ \\
  \hline
  nogrow2e420k\_1	    &   20000 &	-	&   - & -	& -	    & MMSN  & 0.025	& 60	& 50	& 599 & 3$\times$10$^{-4}$ \\
  nogrow5e450k\_1	    &   50000 &	-	&   - & -	& -	    & MMSN	& 0.025	& 130	& 100	& 442 & 3$\times$10$^{-4}$ \\
  \hline
  nogrow2e420k\_hg1	    &   20000 &	-	&   - & -	& -	    & HG	& 0.025	& 60	& 50	& 599 & 3$\times$10$^{-4}$ \\
  nogrow5e450k\_hg1	    &   50000 &	-	&   - & -	& -	    & HG	& 0.025	& 130	& 100	& 442 & 3$\times$10$^{-4}$ \\
  \hline
    \end{tabular}
\end{table*}
\begin{table*}
\centering
\caption{Summary of additional migration simulation parameters.  $N$ is the number of planetesimals in the simulation, $\tau_\mathrm{grow}$ is the growth time, $\tau_\mathrm{mig}$ is the migration timescale, $t_\mathrm{start,J/S}$ is the time at which Jupiter/Saturn begin migrating, $\Delta t$ is the simulation timestep, $\tau_\mathrm{gas}$ is the gas disk dissipation timescale, $R_\mathrm{init}$ is the starting planetesimal radius, and $M_\mathrm{min}$ is the mass resolution limit.
\label{t:sims2}}
\begin{tabular}{lrrrrrrrrrrr}
  \hline
  Name &	$N$ &	$\tau_\mathrm{grow}$	& $\tau_\mathrm{mig}$ 	& $t_\mathrm{start,J}$	& $t_\mathrm{start,S}$	& Disk	& $\Delta t$	& Duration	& $\tau_\mathrm{gas}$ & $R_\mathrm{init}$ & $M_\mathrm{min}$ \\
   & &	(kyr)	& (kyr)		& (kyr)	& (kyr)	&	& (yr)	& (kyr)	& (kyr) & (km) & (\mearth)\\
  \hline
  mig10in5e4\_100k\_1	&   100000	& -	&   50	&       50	&   50	&       MMSN	& 0.05	&   150	&   200 & 465 & 1.33$\times$10$^{-4}$ \\
  mig10in5e4\_100k\_2	&   100000	& -	&   50	&       50	&   50	&       MMSN	& 0.05	&   150	&   200 & 465 & 1.33$\times$10$^{-4}$ \\
  mig10in5e4\_100k\_3	&   100000	& -	&   50	&       50	&   50	&       MMSN	& 0.05	&   150	&   200 & 465 & 1.33$\times$10$^{-4}$ \\
  \hline
  mig10in5e4\_100k\_hg1 &	100000	& -	&   50	&       50	&   50	&       HG	&   0.05	&   150	&   200 & 465 & 1.33$\times$10$^{-4}$ \\
  mig10in5e4\_100k\_hg2	&   100000	& -	&   50	&       50	&   50	&       HG	&   0.05	&   150	&   200 & 465 & 1.33$\times$10$^{-4}$ \\
  mig10in5e4\_100k\_hg3	&   100000	& -	&   50	&       50	&   50	&       HG	&   0.05	&   150	&   200 & 465 & 1.33$\times$10$^{-4}$ \\
  \hline
  mig10in1e5\_100k\_1	&   100000	& -	&   100	&       50	&   50	&       MMSN	& 0.05	&   200	&   200 & 465 & 1.33$\times$10$^{-4}$ \\
  mig10in1e5\_100k\_2	&   100000	& -	&   100	&       50	&   50	&       MMSN	& 0.05	&   200	&   200 & 465 & 1.33$\times$10$^{-4}$ \\
  mig10in1e5\_100k\_3	&   100000	& -	&   100	&       50	&   50	&       MMSN	& 0.05	&   200	&   200 & 465 & 1.33$\times$10$^{-4}$ \\
  \hline
  mig10in1e5\_100k\_hg1	&   100000	& -	&   100	&       50	&   50	&       HG	&   0.05	&   200	&   200 & 465 & 1.33$\times$10$^{-4}$ \\
  mig10in1e5\_100k\_hg2	&   100000	& -	&   100	&       50	&   50	&       HG	&   0.05	&   200	&   200 & 465 & 1.33$\times$10$^{-4}$ \\
  mig10in1e5\_100k\_hg3	&   100000	& -	&   100	&       50	&   50	&       HG	&   0.05	&   200	&   200 & 465 & 1.33$\times$10$^{-4}$ \\
  \hline
  mig10in2e5\_100k\_1	&   100000	& -	&   200	&       50	&   50	&       MMSN	& 0.05	&   300	&   200 & 465 & 1.33$\times$10$^{-4}$ \\
  mig10in2e5\_100k\_2	&   100000	& -	&   200	&       50	&   50	&       MMSN	& 0.05	&   300	&   200 & 465 & 1.33$\times$10$^{-4}$ \\
  mig10in2e5\_100k\_3	&   100000	& -	&   200	&       50	&   50	&       MMSN	& 0.05	&   300	&   200 & 465 & 1.33$\times$10$^{-4}$ \\
  \hline
    \end{tabular}
\end{table*}
\begin{table*}
\centering
\caption{Summary of additional Grand Tack simulation parameters. Those with the tack at 1.5\,au have `GT15' in their names, those with the tack at 2.0\,au instead have names beginning `GT20'.  $N$ is the number of planetesimals in the simulation, $\tau_\mathrm{grow}$ is the growth time, $\tau_\mathrm{mig}$ gives the migration timescales for Jupiter and Saturn, $t_\mathrm{start,J/S}$ give the times at which Jupiter/Saturn begin growth and migration, $\Delta t$ is the simulation timestep, $\tau_\mathrm{gas}$ is the gas disk dissipation timescale, $R_\mathrm{init}$ is the starting planetesimal radius, and $M_\mathrm{min}$ is the mass resolution limit.
\label{t:sims3}}
\begin{tabular}{lrrrrrrrrrrr}
  \hline
  Name &	$N$ &	$\tau_\mathrm{grow}$	& $\tau_\mathrm{mig}$ 	& $t_\mathrm{start,J}$	& $t_\mathrm{start,S}$	& Disk	& $\Delta t$	& Duration	& $\tau_\mathrm{gas}$ & $R_\mathrm{init}$ & $M_\mathrm{min}$ \\
   & &	(kyr)	& (kyr)		& (kyr)	& (kyr)	&	& (yr)	& (kyr)	& (kyr) & (km) & (\mearth)\\
  \hline
  GT15gm5e4\_50k\_1	&       50000	& 50	& 50, 5	&   0, 50	& 50, 95	& MMSN	& 0.0167	& 250 & 100	& 571 & 2.5$\times$10$^{-4}$ \\
  GT15gm5e4\_100k\_1	&   100000	& 50    & 50, 5   & 0, 50   & 50, 95    & MMSN	& 0.0167    & 200 & 100 &  453 & 1.17$\times$10$^{-4}$ \\
  \hline
  GT15gm5e4\_50k\_hg1 &     50000	& 50	& 50, 5	&   0, 50	& 50, 95	& HG	& 0.0167	& 250 & 100	  & 571 & 2.5$\times$10$^{-4}$ \\
  GT15gm5e4\_50k\_hg2 &     50000	& 50	& 50, 5	&   0, 50	& 50, 95	& HG	& 0.0167	& 250 & 100	  & 571 & 2.5$\times$10$^{-4}$ \\
  GT15gm5e4\_100k\_hg1	&   100000	& 50    & 50, 5   & 0, 50   & 50, 95    & HG	& 0.0167    & 200 & 100	& 453 & 1.17$\times$10$^{-4}$ \\
  \hline
  GT20gm5e4\_50k\_hg1 &     50000	& 50	& 50, 5	&   0, 50	& 50, 95	& HG	& 0.0167	& 250 & 100	& 571 & 2.5$\times$10$^{-4}$ \\
  GT20gm5e4\_50k\_hg2 &     50000	& 50	& 50, 5	&   0, 50	& 50, 95	& HG	& 0.0167	& 250 & 100	& 571 & 2.5$\times$10$^{-4}$ \\
  \hline
  GT15gm1e5\_50k\_1	&       50000	& 100	& 100, 10 & 0, 100	& 100, 190	& MMSN	& 0.0167	& 500 & 100	& 571 & 2.5$\times$10$^{-4}$ \\
  \hline
  GT15gm1e5\_50k\_hg1 &     50000	& 100	& 100, 10 & 0, 100	& 100, 190	& HG	& 0.0167	& 500 & 100	& 571 & 2.5$\times$10$^{-4}$ \\
  \hline
  GT20gm1e5\_50k\_hg1 &     50000	& 100	& 100, 10 & 0, 100	& 100, 190	& HG	& 0.0167	& 500 & 100	& 571 & 2.5$\times$10$^{-4}$ \\
  \hline
    \end{tabular}
\end{table*}
%


\end{document}